# Variational and Monte Carlo Methods for Bayesian Inversion of Dynamic Subsurface Flow Simulations Using Seismic and Fluid Pressure Data

**Zhen Zhang[1], Xuebin Zhao[1], and Andrew Curtis[1]**

[1]University of Edinburgh, School of Geosciences, EH9 3FE, Edinburgh, U.K.

Corresponding author: Zhen Zhang (s2673720@ed.ac.uk); Andrew Curtis (Andrew.Curtis@ed.ac.uk)

**Key Points:**

- Variational inference (VI) methods impose specific parametrisations on Bayesian posterior probability distributions, greatly improves which the computational efficiency of uncertainty estimation.
- Automatic differentiation VI methods are extremely efficient, while sample-based Stein variational gradient descent methods may offer more accurate uncertainty estimates at far higher computational cost.

**Abstract**

In order to predict future performance of subsurface fluid reservoirs under possible operating scenarios, a dynamic, porous-medium flow simulation model must be tuned to include representative properties of the reservoir. Estimating subsurface reservoir properties given remotely sensed or borehole-based observations typically involves finding the solution to a challenging inverse problem. We compare Monte Carlo random sampling to variational inference methods which use optimisation to constrain parametrised uncertainties in nonlinear Bayesian inversions. We use them to estimate the posterior probability distribution of reservoir permeability given fluid pressure and seismic measurements. The methods include automatic differentiation variational inference (ADVI), Stein variational gradient descent (SVGD), and a Monte Carlo method called stochastic SVGD (sSVGD), all of which we benchmark against results from Metropolis-Hastings McMC. We also test an ADVI variant called physically structured variational inference (PSVI): in our implementation this method estimates only spatially-local correlations between model parameters based on the intuition that such correlations are strong in remote sensing problems in which data only inform about spatial-averages of local dynamics. We apply the methods to two- and three-dimensional inverse problems of carbon dioxide storage, inspired by the Endurance field, located in the UK North Sea. Results show that PSVI achieves a good balance between mean-field ADVI and full-rank ADVI in terms of accuracy of the posterior approximation and computational efficiency. SVGD and sSVGD offer more accurate approximations of the target posterior distribution, but at far higher computational cost. Between them, sSVGD outperforms SVGD, exhibiting better computational efficiency and mitigating the problems of mode collapse and spurious correlations.

**Plain Language Summary**

Predicting how subsurface fluid reservoirs will behave in the future requires flow simulation models that use unknown properties of the reservoir. These properties cannot be measured directly, so they must be estimated from indirect observations such as borehole-based pressure data or seismic measurements. Their estimation is difficult because many different property combinations can explain the same observations. Bayesian methods address these issues by estimating the range of possible property values and their probabilities. Random sampling Markov chain Monte Carlo (McMC) methods can do this accurately but become too slow when the number of unknowns is large. We therefore compare McMC with faster variational inference approaches that optimise an approximation to the range of possibilities. These include automatic differentiation variational inference (ADVI), Stein variational gradient descent (SVGD), stochastic SVGD (sSVGD), and a new ADVI variant, physically structured variational inference (PSVI) which focuses on trade-offs between spatially-neighbouring parameters which are thought to dominate in such problems. We apply these methods to two- and three-dimensional inverse problems of carbon dioxide storage, inspired by the Endurance field, located in the UK North Sea. PSVI provides a good balance between computational efficiency and accuracy, while SVGD and sSVGD captures complex uncertainties more faithfully but at higher cost.

## 1 Introduction

Predictive models are essential to inform decision-making in a range of applications involving fluid flow in the Earth's subsurface, including hydrological modelling, $CO_2$ sequestration, and geothermal energy extraction amongst others (Driesner & Geiger, 2007; Li et al., 2020; Singh et al., 2010; Sifuentes et al., 2009). If we wish to use computational models to predict future responses of dynamical systems with meaningful associated uncertainties, it is generally necessary to estimate physically representative values of the controlling model parameters (Arnold et al., 2013; Ibiam et al., 2021). This study addresses how such representative values can be estimated.

Inferring the values of model parameters is often challenging, and since fluid flow simulators (here referred to as *forward* models) generally predict flow given parameter values but not parameters from flow observations, the latter involves solving an inverse problem. The variety of existing inversion methods can broadly be categorized into two types: deterministic and probabilistic approaches (Xia & Zabaras, 2022). Deterministic inversion involves finding the model parameter values that minimise the discrepancy between model predictions and field observations. Various optimisation algorithms have been proposed for this purpose, including iterated locally-linearised optimisation, and global methods that sample a broader range of parameter values such as genetic algorithms, particle swarm optimisation, and simulated annealing (Maschio et al., 2008; Maschio & Schiozer, 2018; Rawlinson et al., 2010; Xavier et al., 2013; Yin et al., 2020).

Locally-linearised approaches compute the gradient of an objective function (a measure of discrepancy between observed data and data simulated from the model) with respect to the model parameters, allowing parameters to be perturbed iteratively so as to reduce the misfit. Commonly used examples include the Levenberg–Marquardt and quasi-Newton methods (Bjarkason et al., 2018; Eltahan et al., 2024; Kaleta et al., 2011). Regularisation must be incorporated to ensure that the methods converge, and that the results exhibit some chosen characteristics. This is typically achieved by adding spatial or temporal smoothing or perturbation-damping terms to the objective function, or simply by terminating iterations before perturbations from reference characteristics become too large (Brun et al., 2004;

Golmohammadi et al., 2015; Waldron et al., 2020). While regularisation stabilizes the otherwise ill-posed and under-constrained inversions, and reduces overfitting of the data (finding parameter values that fit components of the measurements caused by observational errors), it usually also introduces biases into the final solution by favouring parameter combinations that deviate from more realistic values.

A second, key limitation of local optimisation methods is their inability to account for uncertainties in model parameters. In fluid flow inverse problems, the strong nonlinearity of the model-data relationship and the inherent non-uniqueness of these ill-posed problems mean that an infinite number of sets of model parameter values produce equally good fits to observations (Oliver & Chen, 2011). It is therefore necessary to estimate or bound the family of parameter value combinations that are consistent with observations (the uncertainty). While the global optimisation methods listed above explore more of parameter space than local methods, they are designed for optimisation so generally do not provide samples that are suitable for estimating uncertainties robustly.

Bayesian inversion (often called Bayesian inference) provides a rigorous framework for estimating uncertainties described by a probability density function (pdf) over the model parameter values, known as the posterior distribution. Obtaining a closed-form expression for the posterior solution is usually infeasible because nonlinearity in the forward function renders analytical mathematical methods ineffective. Therefore, sampling-based approaches such as Markov chain Monte Carlo (McMC) are commonly used to *explore* the posterior distribution (Hastings, 1970; Metropolis et al., 1953). A perceived advantage of McMC is its asymptotic accuracy: given an infinite number of samples, it can accurately sample the target posterior distribution. As a result, McMC methods have been applied widely across various disciplines for parameter estimation: for example, Mosegaard & Tarantola (1995) first applied McMC to geophysical problems in a gravity data inversion, Malinverno (2002) inverted direct current resistivity data to characterise electrical conductivity in layered earth structures, Chen et al. (2014) inverted for thermal conductivity and formation permeability in a geothermal reservoir, Tripoppoom et al. (2020) inferred hydraulic fracture properties from production data in a fractured reservoir, and Zhang et al. (2023) estimated fluid viscosity and permeability using production data from an oil field. Efendiev et al. (2009) proposed a two stage McMC in which the first stage is run in coarse scale sampled simulation model to determine whether to accept it or not; if the coarse mode is accepted, a finer scale simulation model is run to update the model parameters, while another coarse scale model is proposed for the next iteration. This reduces the computation time by avoiding running a large number of finer simulation models. However, a significant limitation of McMC is its computational requirements in high-dimensional inverse problems due to the so-called curse of dimensionality: as the number of model parameters increases, the number of samples required for statistical convergence to the posterior pdf grows exponentially (Curtis & Lomax, 2001). It is therefore likely that asymptotic accuracy is never achieved in practical, high-dimensional flow problems.

To address this challenge, Kalman filter-based methods compute parameter updates that increase information when assimilating newly observed data, in a way that amounts to applying Bayes rule in linearised problems (Bruer et al., 2025; Evensen, 2009). Uncertainties are represented by a set of samples (called particles) in parameter space, and under certain conditions these approaches require substantially fewer forward simulations than McMC and have demonstrated effective estimation of uncertainties across a range of applications. For example, Chen et al. (2013) applied the ensemble Kalman filter (EnKF) to invert for permeability models using tracer data, Gharamti et al. (2014) used EnKF to estimate permeability in compositional flow models from flow-chemical data, and Emerick & Reynolds (2013) employed the ensemble smoother with multiple data assimilation (ES-MDA) to estimate permeability from production data. Tavakoli et al. (2013) compared the performance of several Kalman filter–based methods for estimating permeability in a $CO_2$ sequestration problem, including the EnKF, ensemble smoother (ES), and ES-MDA, and found that ES-MDA achieved slightly better computational efficiency and accuracy than the other two methods. Although Kalman filter-based methods can be more computationally efficient than McMC, they have notable limitations: in particular, in order to keep track of uncertainty correctly they require that the relationships between model parameters and system responses are either linear, or are nonlinear in a way that only one connected, convex region of high probability exists in parameter space at each step – neither of which seem likely to be true in complex subsurface flow problems. Moreover, even in pathological cases in which these conditions are met, the ensemble size must increase with the dimensionality of the model parameters to ensure reliable covariance estimation during assimilation, which again leads to high computational costs in high-dimensional inverse problems.

To reduce computational cost, various machine learning based methods have been proposed to assist in estimating model parameters. One common approach is to train a machine learning model to serve as a surrogate for the computationally expensive forward simulations. For example, Maschio & Schiozer (2014) employed artificial neural networks (ANNs) to replace high-cost forward simulators, and then applied McMC to invert porosity and permeability. Similarly, Tripoppoom et al. (2020) combined K-nearest neighbours and ANNs to substitute for forward flow simulations in the estimation of fracture properties using McMC. Wang et al. (2022) integrated theory-guided neural networks with an iterative ensemble smoother to estimate reservoir permeability. An advantage of these approaches is that the forward function is usually unique-valued, so the output of the surrogate function may also be so. However, in order to train such surrogates, many examples of the forward function applied to model samples must first be evaluated explicitly (Yin et al., 2022), and practical experience in testing these methods commonly produces examples in which training set sizes turn out to be insufficient, leading to instability and errors in surrogate outputs (e.g., Krishnapriyan et al., 2021; McGreivy and Hakim, 2024). This in turn leads to unreliable inverse problem uncertainty estimates, and we note that it also seems likely that most such cases are never published, leading to a positivity bias in existing literature (Duyx et al., 2017).

Another potential approach is to train a neural network to model the inverse relationship from data to model parameters directly. In this framework, once the machine learning model has been trained, model parameters can be estimated from new observations rapidly. For example, Earp & Curtis (2020) and Zhang & Curtis (2021), respectively, introduced mixture density and invertible networks to perform probabilistic seismic tomography given travel time data, Ishitsuka & Lin (2023) proposed a physics-informed neural network (PINN) to predict spatial permeability, temperature and pressure using sparse well observations with the PINN trained to satisfy the conservation of mass and energy, and Suzuki et al. (2024) developed a machine-learning based surrogate model to directly map reservoir pressure and temperature to permeability.

Although such machine learning methods can be highly efficient once trained, the training process itself is usually computationally expensive. Generating training data involves generating a representative set of samples from the *prior* distribution of model parameter values (the prior distribution describes the state of knowledge about parameter values a *priori*-before current data have been considered). The forward simulation must then be run for each sample to obtain corresponding model outputs. Since the portion of the prior distribution with probability significantly greater than zero (called the support of the distribution) almost always spans a far larger subset of parameter space than does the equivalent for the posterior distribution, the sample set must be correspondingly far larger to explore the prior than to explore the posterior pdf, leading to a substantially increased number of simulations. Moreover, these machine learning inversion methods are often problem-specific, meaning that any change in the model or data acquisition configuration generally requires that the machine learning model is retrained (Wen et al., 2021), although there are exceptions (Zhang et al., 2025).

To mitigate the aforementioned limitations, we introduce variational inference to solve fluid flow related inverse problems. Unlike random-sampling based approaches such as Markov chain Monte Carlo (McMC), variational inference reformulates the problem as an optimisation. It approximates the target posterior pdf over model parameters by warping a predefined probability distribution to minimise its deviation from the target. Although this approach is conceptually similar to machine learning methods that minimise objective functions, it differs in several important ways. First, we do not need to train a surrogate model to replace the forward simulator, nor to learn a mapping from all possible observations to model parameters (although both of these can be incorporated if desired-Ganguly et al., 2023; Kingma & Welling, 2014). The variational methods that we introduce here optimise a predefined form of distribution to best approximate the target posterior distribution for a single dataset, avoiding the need to explore the entire support of the prior pdf. Furthermore, similarly to Markoc chain Monte Carlo, variational methods avoid the need to compute a particular term in Bayes' rule called the evidence (see below), which would otherwise require an exploration of the full parameter space which is intractable in large numerical inverse problems (Zhao et al., 2022). While constraining the problem to finding the best member of a specific class of (so-called *variational*) probability distributions produces only an approximation to the posterior pdf, so do all sampling based methods in practice due to the curse of dimensionality, and it is this approximation that makes variational methods efficient. Defining different classes of variational distributions leads to different algorithms.

In previous studies, a commonly chosen variational inference method is Stein variational gradient descent (SVGD). This method iteratively refines a set of samples (called particles in SVGD because they 'move') from their initial values or positions in parameter space toward positions for which their joint density better approximates the posterior distribution (Liu & Wang, 2016). By using the Jacobian gradients of data with respect to parameter values to progressively reduce the discrepancy between the posterior target and the particle-based density approximation, SVGD estimates the posterior pdf as an optimisation problem, avoiding the intensive random sampling demands required by McMC in some classes of problems. SVGD has shown promising performance in geophysical applications including full waveform inversion (Zhang & Curtis, 2020b) and travel time tomography (Zhang & Curtis, 2020a). Inin hydrogeological modelling, Ramgraber et al. (2021) used SVGD to estimate uncertain parameters in a hydrogeological model using a kernel-based Jacobian approximation to reduce the computational cost of the Jacobian evaluations. More recently, Liu et al. (2024) used SVGD to approximate the posterior pdf of clay volume and porosity parameters in a pre-stack seismic inversion. An autoencoder (AE) was employed to reduce the dimensionality of the model parameters, and they showed that AE-assisted SVGD can approximate complex posterior pdfs with improved computational efficiency. Peng et al. (2025) trained a Bayesian neural network using SVGD to characterise the posterior distribution of the network parameters, enabling uncertainty quantification in groundwater level predictions.

Despite these advantages, SVGD requires many particles to sufficiently approximate a posterior distribution in high-dimensional space: only the sample density around each point in parameter space represents that distribution, and many samples (particles) are needed around any point to estimate a reliable density. With insufficient particles, SVGD can suffer from mode collapse, where most samples concentrate in a small region and fail to represent the full posterior distribution adequately (Angelo & Fortuin, 2021; Zhang et al., 2020). A variation of the SVGD method known as stochastic SVGD (sSVGD) is derived from the solution to a particular stochastic differential equation. In practice, this method discretises parameter space and introduces a random noise term to the SVGD objective function at each iteration (Gallego & Insua, 2018). This stochastic component ensures that the algorithm converges asymptotically to the target posterior pdf of the model parameters, transforming SVGD into a variant of McMC which takes advantage of Stein variational gradient operators to accelerate convergence. Samples evaluated as the particles evolve then also contribute to the solution, greatly increasing the number contributing to the final density estimate. The sSVGD approach has previously been shown to be effective in high-dimensional full waveform inversion (Zhang et al., 2023; Zhang and Curtis, 2024).

Automatic differentiation variational inference (ADVI) is another widely used variational inference method (Kucukelbir et al., 2017), where a predefined family of variational distributions is optimised to approximate the target posterior pdf. Various probability density functions (pdfs) have been employed as variational families for this purpose. Among them, the Gaussian distribution has demonstrated strong performance across a range of applications. For example, Krapu et al. (2019) used a Gaussian variational family within ADVI to approximate the posterior distribution of 25 years of daily precipitation in a rainfall-runoff model, conditioned on streamflow observations. Li et al. (2021) employed a Gaussian variational family to approximate the posterior distribution of the parameters in a Bayesian neural network, aiming to improve predictive uncertainty in process-based hydrological models. Zhang & Curtis (2020a) applied a Gaussian distribution to approximate the posterior pdf of the seismic velocity field in travel time tomography, and Zhang & Curtis (2020b, 2024) did the same for full waveform inversion. Zhao & Curtis (2024a) then introduced a Gaussian mixture model (a sum of multiple Gaussians) to more reliably capture complex, multimodal posterior distributions in velocity estimation for full waveform inversion. Despite such extensions, the single Gaussian distribution remains a popular choice in practice due to its favourable balance between accuracy and computational efficiency.

Finding a Gaussian approximation to the posterior pdf using ADVI requires the estimation of an appropriate mean vector and covariance matrix. Different choices of the freedoms allowed within the covariance matrix result in different variants of the method. For example, the mean field approximation assumes a factorised (uncorrelated) Gaussian distribution by using a diagonal covariance matrix, significantly simplifying computations. However, this assumption neglects important spatial correlations in subsurface properties, leading to underestimation of uncertainties in posterior parameter marginal distributions (Zhang et al., 2023). This may be particularly important in the estimation of subsurface reservoir properties, in which spatial correlations play a critical role. Full-rank ADVI relaxes the assumption of parameter independence by allowing a dense covariance matrix, which can improve the representation of posterior dependencies and the marginal uncertainties of the model parameters. However, this comes at a substantial computational cost and memory requirement: in high-dimensional inverse problems, storing and optimising a full-rank covariance matrix becomes impractical, severely limiting the scalability of full-rank ADVI in geophysical inverse problems (Zhang et al., 2023; Zhao and Curtis, 2025a, 2024a).

Mean-field ADVI and full-rank ADVI represent two extreme situations in the representation of the covariance matrix, suffering from limited correlation modelling and high computational cost, respectively. To balance accuracy and efficiency, physically structured variational inference (PSVI) uses a sparse representation of the covariance matrix which only captures spatial correlations that are expected to be substantial *a priori*, given knowledge of the physics of each particular inverse problem (Zhao & Curtis, 2024b). The method has been applied successfully to high-dimensional geophysical inversion problems with computationally demanding forward problems such as 3D full waveform inversion (Zhao & Curtis, 2025b).

In fluid flow applications data are recorded at locations that are remote from the locations of most reservoir parameters to be estimated, and the physics of fluid flow mixes or confounds information about medium properties at previous locations carried by packets of fluid as they travel through a porous medium (Rey et al., 2012). An inverse problem set up to estimate those medium properties is therefore only expected to offer limited spatial resolution aound each parameter location. This means that only locally spatially-averaged flow properties are constrained by the data, so that correlations between parameter estimates will be high between neighbouring locations below some spatial scale. This characteristic motivates the use of inference methods that can capture at least local correlations, such as PSVI, full-rank ADVI, SVGD or sSVGD.

As a result of the above discussion, this study assesses the effectiveness of these particular variational inference and Monte Carlo methods in both 2D and 3D $CO_2$ sequestration fluid flow problems. The paper is organized as follows: Section 2 provides a detailed overview of variational inference, with an emphasis on ADVI and its limitations, and also introduces the PSVI method in more detail. Section 3 assesses the performance of ADVI, PSVI, SVGD, sSVGD and a standard Metropolis-Hastings McMC on both the 2D and 3D problems. Section 4 discusses implications of the results, and Section 5 summarises the main conclusions from this work.

## 2 Methods

The objective of this work is to characterise the posterior pdf $p(\mathbf{m}|\mathbf{d}_{obs})$, which describes the state of information about model parameter $\mathbf{m}$ given observed data $\mathbf{d}_{obs}$. According to Bayes' theorem, the posterior pdf is given by:

$$p(\mathbf{m}|\mathbf{d}_{obs}) = \frac{p(\mathbf{d}_{obs}|\mathbf{m})p(\mathbf{m})}{p(\mathbf{d}_{obs})} \tag{1}$$

where $p(\mathbf{d}_{obs}|\mathbf{m})$ denotes the likelihood which represents the relative probability of observing data $\mathbf{d}_{obs}$ given a particular set of parameter values $\mathbf{m}$, $p(\mathbf{m})$ is the prior distribution which reflects our knowledge about the parameters before incorporating any observations, and $p(\mathbf{d}_{obs})$ is the evidence which in this case serves as a normalisation constant to ensure that the posterior distribution is a valid pdf.

### 2.1 Variational Inference

Variational inference methods approximate the target posterior pdf $p(\mathbf{m}|\mathbf{d}_{obs})$ of the model parameter $\mathbf{m}$ by a simpler, parameterised distribution $q(\mathbf{m})$ from a predefined family of distributions. The goal is to find an optimal distribution $q(\mathbf{m})$ that minimises the Kullback-Leibler (KL) divergence (Kullback and Leibler, 1951) between $q(\mathbf{m})$ and the target posterior pdf $p(\mathbf{m}|\mathbf{d}_{obs})$:

$$\mathrm{KL}\big(q(\mathbf{m})|p(\mathbf{m}|\mathbf{d}_{obs})\big) = E_q[\log q(\mathbf{m})] - E_q[\log p(\mathbf{m}|\mathbf{d}_{obs})] \tag{2}$$

where $E_q$ is the expectation with respect to $q(\mathbf{m})$, and the KL divergence is a measure of difference between two different distributions. The KL divergence is always non-negative and equals zero only when $q(\mathbf{m}) = p(\mathbf{m}|\mathbf{d}_{obs})$ (Kullback and Leibler, 1951). Substituting Bayes' theorem into equation (2), we obtain

$$\mathrm{KL}\big(q(\mathbf{m})|p(\mathbf{m}|\mathbf{d}_{obs})\big) = E_q[\log q(\mathbf{m})] - E_q\left[\log\frac{p(\mathbf{d}_{obs}|\mathbf{m})p(\mathbf{m})}{p(\mathbf{d}_{obs})}\right] \quad (3)$$

Define $p(\mathbf{m}, \mathbf{d}_{obs})$ is the joint probability distribution over $\mathbf{m}$ and $\mathbf{d}_{obs}$

$$p(\mathbf{m}, \mathbf{d}_{obs}) = p(\mathbf{d}_{obs}|\mathbf{m})p(\mathbf{m}) \quad (4)$$

we then obtain

$$\mathrm{KL}\big(q(\mathbf{m})|p(\mathbf{m}|\mathbf{d}_{obs})\big) = E_q[\log q(\mathbf{m})] - E_q[\log p(\mathbf{m}, \mathbf{d}_{obs})] + \log p(\mathbf{d}_{obs}) \quad (5)$$

The evidence term $\log p(\mathbf{d}_{obs})$ involves a high-dimensional integral that is often computationally intractable. However, since this term is constant, minimising the KL divergence is equivalent to minimising the combination of the other two terms, or maximising the so-called evidence lower bound (ELBO) defined by rearranging the KL divergence as follows:

$$\begin{aligned}\mathrm{ELBO} &= \log p(\mathbf{d}_{obs}) - \mathrm{KL}\big(q(\mathbf{m})|p(\mathbf{m}|\mathbf{d}_{obs})\big) \\ &= E_q[\log p(\mathbf{m}, \mathbf{d}_{obs})] - E_q[\log q(\mathbf{m})]\end{aligned} \quad (6)$$

Since $\mathrm{KL}\big(q(\mathbf{m})|p(\mathbf{m}|\mathbf{d}_{obs})\big) \geq 0$, maximising the ELBO is equivalent to minimising the KL divergence. The ELBO is computationally tractable and serves as the objective function in most variational inference methods.

The choice of variational family defined by the set of possible distributions $q(\mathbf{m})$ plays a critical role in the accuracy of the posterior pdf approximation. A more expressive distribution can better capture the characteristics of the posterior pdf, but often requires higher computational cost to achieve a reliable optimisation result (Gahlot et al., 2025; Zhao et al., 2022). Conversely, simpler distributions offer greater computational efficiency but may fail to represent key features of the posterior distribution. Herein we test several methods that in principle may be capable of representing highly complex distributions (SVGD, sSVGD and Metropolis-Hastings Monte Carlo).

However, in order to balance expressiveness and efficiency, with most focus on efficiency since we wish to apply these methods to high-dimensional problems, we also test the use of a Gaussian distributions of various complexities to approximate the posterior pdf of the model parameters in the following class of methods.

## 2.2 Automatic Differentiation Variational Inference (ADVI)

Kucukelbir et al. (2017) proposed a variational inference method using a Gaussian variational family called automatic differentiation variational inference (ADVI). ADVI aims to approximate the target posterior pdf as shown in Figure 1a. Since the posterior distribution of the parameters is often defined over a constrained space, for example permeability values are non-negative and typically bounded, to allow the variational approximation to be chosen independently of the original bounds we first transform the constrained space into an unconstrained (infinitely wide) space (Figure 1b). This transformation allows the infinitely wide Gaussian variational family to be optimised to fit the transformed posterior pdf, by maximising the evidence lower bound (ELBO). After optimisation, the fitted Gaussian distribution is transformed back into the constrained space to characterise the posterior distribution within the original parameterisation.

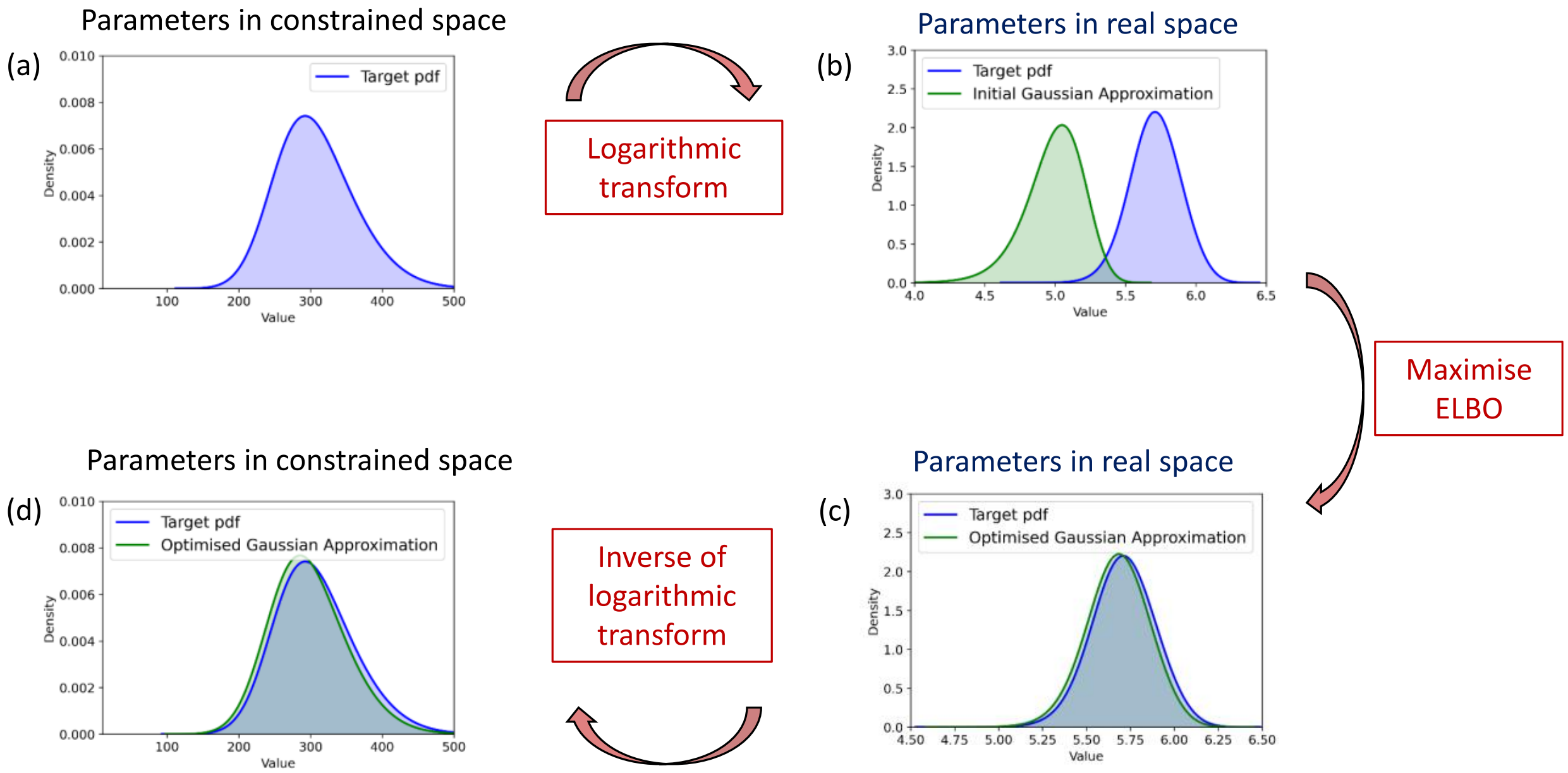

**Figure 1.** Workflow of ADVI. (a) The unknown target posterior pdf (blue) that we wish to estimate. (b) The parameter value is transformed to an unconstrained space and an initial variational Gaussian distribution is used to approximate the target pdf; the target distribution is also represented but remains unknown throughout. (c) The Gaussian approximation to the target posterior pdf is optimised by maximising the ELBO. (d) The optimised Gaussian distribution is mapped back to the constrained space to produce a final variational solution.

To transform parameter $\mathbf{m}$ from the constrained space, such as a physical space that has bounds of 0 and 1 on porosity, into a space of real numbers, we use a differentiable and invertible function $\boldsymbol{\mathcal{S}} = T(\mathbf{m})$. A common choice for this transformation is the logistic function (Kucukelbir et al., 2017; Zhang & Curtis, 2020a)

$$\mathcal{S}_i = T(m_i) = \log(m_i - a_i) - \log(b_i - m_i)$$
$$m_i = T^{-1}(\mathcal{S}_i) = a_i + \frac{b_i - a_i}{1 + \exp(-\mathcal{S}_i)} \tag{7}$$

where $m_i$ and $\mathcal{S}_i$ are the parameters in the original and transformed space, respectively, and $a_i$ and $b_i$ are the lower and upper bound on $m_i$. During this transformation, it is essential to account for the Jacobian matrix of the inverse transformation $T^{-1}$, denoted as $\mathbf{J}_{T^{-1}}(\boldsymbol{\mathcal{S}})$, which quantifies local volumetric changes between the spaces and in turn affects probability densities. Therefore, the probability density function $p(\boldsymbol{\mathcal{S}}, \mathbf{d}_{obs})$ in the transformed space is adjusted to account for the Jacobian of the inverse transformation:

$$p(\boldsymbol{\mathcal{S}}, \mathbf{d}_{obs}) = p(\mathbf{m}, \mathbf{d}_{obs})|\det \mathbf{J}_{T^{-1}}(\boldsymbol{\mathcal{S}})| \tag{8}$$

In the unconstrained space, the variational distribution is defined by

$$q(\boldsymbol{\mathcal{S}}; \boldsymbol{\varphi}) = N(\boldsymbol{\mathcal{S}}|\boldsymbol{\mu}, \boldsymbol{\Sigma}) = N(\boldsymbol{\mathcal{S}}|\boldsymbol{\mu}, \mathbf{L}\mathbf{L}^{\mathbf{T}}) \tag{9}$$

where $\boldsymbol{\varphi}$ denotes the variational parameter, which includes the Gaussian mean vector $\boldsymbol{\mu}$ and the covariance matrix $\boldsymbol{\Sigma} = \mathbf{L}\mathbf{L}^{\mathbf{T}}$, with $\mathbf{L}$ being a lower triangular matrix. $N$ is standard notation for a Gaussian (normal) distribution. The matrix $\mathbf{L}$ is derived from the Cholesky decomposition of $\boldsymbol{\Sigma}$. This decomposition simplifies the computation of the covariance matrix determinant because $|\boldsymbol{\Sigma}| = |\mathbf{L}|^2$, and the determinant of $\mathbf{L}$ is simply the product of its diagonal elements.

As shown in Figure 1c, the goal is to maximise the ELBO to optimise the variational parameter $\boldsymbol{\varphi}$:

$$\boldsymbol{\varphi}^* = \mathrm{argmax}_{\boldsymbol{\varphi}}\ \mathrm{ELBO}\ [q(\boldsymbol{\mathcal{S}}; \boldsymbol{\varphi})] \tag{10}$$

If we substitute

$$\text{ELBO}\,[q(\boldsymbol{\mathcal{S}};\boldsymbol{\varphi})] = E_q[\log p(T^{-1}(\boldsymbol{\mathcal{S}}),\mathbf{d}_{obs})] + \log|\det \mathbf{J}_{T^{-1}}(\boldsymbol{\mathcal{S}})| \; - E_q[\log q(\boldsymbol{\mathcal{S}};\boldsymbol{\varphi})] \tag{11}$$

into equation (10) we obtain

$$\boldsymbol{\varphi}^* = \text{argmax}_{\boldsymbol{\varphi}}\; E_q[\log p(T^{-1}(\boldsymbol{\mathcal{S}}),\mathbf{d}_{obs})] + \log|\det \mathbf{J}_{T^{-1}}(\boldsymbol{\mathcal{S}})| \; - E_q[\log q(\boldsymbol{\mathcal{S}})] \tag{12}$$

The optimisation problem in Equation (12) can be solved efficiently using gradient ascent methods. To obtain unbiased estimates of the ELBO and its gradients when using Monte Carlo sampling, we apply a reparameterisation that transforms the variational distribution $q(\boldsymbol{\mathcal{S}};\boldsymbol{\varphi})$ into a standard Gaussian distribution $N(\boldsymbol{\eta}|\mathbf{0},\mathbf{I})$ via the transformation $\boldsymbol{\eta} = R_{\boldsymbol{\varphi}}(\boldsymbol{\mathcal{S}}) = \mathbf{L}^{-1}(\boldsymbol{\mathcal{S}} - \boldsymbol{\mu})$. Thereafter, equation (12) can be reformulated as

$$\begin{aligned}\boldsymbol{\varphi}^* = \text{argmax}_{\boldsymbol{\varphi}} E_{N(\boldsymbol{\eta}|\mathbf{0},\mathbf{I})}\left[\log p\left(T^{-1}\left(R_{\boldsymbol{\varphi}}^{-1}(\boldsymbol{\eta})\right),\mathbf{d}_{obs}\right)\right] \\ + \log\left|\det \mathbf{J}_{T^{-1}}\left(R_{\boldsymbol{\varphi}}^{-1}(\boldsymbol{\eta})\right)\right| - E_q[\log q(\boldsymbol{\mathcal{S}};\boldsymbol{\varphi})]\end{aligned} \tag{13}$$

Here, the expectation in the first term is now taken over a standard Gaussian distribution, which simplifies the gradient calculation of the ELBO as follows. By expressing the expectation independently of the variational parameters, the Dominated Convergence Theorem (Cınlar, 2011) allows us to interchange the expectation and derivatives (see Appendix A): applying the chain rule then yields

$$\nabla_{\boldsymbol{\mu}}\text{ELBO} = E_{N(\boldsymbol{\eta}|\mathbf{0},\mathbf{I})}[\nabla_{\mathbf{m}}\log p(\mathbf{m},\mathbf{d}_{obs})\nabla_{\boldsymbol{\mathcal{S}}}T^{-1}(\boldsymbol{\mathcal{S}}) + \nabla_{\boldsymbol{\mathcal{S}}}\log|\det \mathbf{J}_{T^{-1}}(\boldsymbol{\mathcal{S}})|] \tag{14}$$

$$\nabla_{\mathbf{L}}\text{ELBO} = E_{N(\boldsymbol{\eta}|\mathbf{0},\mathbf{I})}[\nabla_{\mathbf{m}}\log p(\mathbf{m},\mathbf{d}_{obs})\nabla_{\boldsymbol{\mathcal{S}}}T^{-1}(\boldsymbol{\mathcal{S}}) + \nabla_{\boldsymbol{\mathcal{S}}}\log|\det \mathbf{J}_{T^{-1}}(\boldsymbol{\mathcal{S}})|\,\boldsymbol{\eta}^T] + (\mathbf{L}^{-1})^T \tag{15}$$

(Kucukelbir et al., 2017). To compute these expectations with respect to the standard Gaussian distribution, Monte Carlo integration is typically employed. While using a larger number of samples generally enhances the precision of these estimates, practical scenarios often allow for the use of fewer samples. This efficiency arises because the iterative nature of the optimisation process means that over many iterations, the use of a limited sample size per iteration can still involve a large number of samples overall and thus yield a stable final variational distribution, as is common in other stochastic optimisation algorithms (Nemirovski et al., 2008; Zhang & Curtis, 2020a).

After optimisation, the variational distribution provides an approximation of the posterior distribution in the unconstrained space, as shown in Figure 1c. The optimised distribution is then transformed back to the original constrained space, shown in Figure 1d, to approximate the posterior pdf of the model parameters in their original domain.

One important aspect to highlight is the choice of covariance matrix in ADVI, as introduced above. A common approach is to optimise a parameterised full-rank covariance matrix, known as full-rank ADVI. This method optimises all elements of the Cholesky factor $\mathbf{L}$, capturing correlations among all model parameter $\mathbf{m}$. However, this results in a large number of variational parameters to store and optimise, at substantial computational cost. Alternatively, the mean-field ADVI method assumes independence among parameters by ignoring parameter correlations explicitly. Therefore, the covariance matrix becomes a diagonal matrix. This simplification greatly reduces the number of parameters to be optimised during inversion, making optimisation more efficient. Yet, it may overlook important correlations, often leading to an underestimation of posterior uncertainty.

## 2.3 The Physically Structured Variational Inference (PSVI)

The physically structured variational inference (PSVI) method parameterises variational distributions, in this Gaussian case the matrix $\mathbf{L}$, using a physics-guided approach (Zhao & Curtis, 2024b). Instead of modelling all off-diagonal elements, in PSVI we selectively include the most significant correlations among model parameters, as expected *a priori* from physics. This approach enables the incorporation of key, known dependencies among model parameters while maintaining a lower computational cost than full-rank ADVI.

To achieve this, a sparse structure is constructed for the matrix $\mathbf{L}$, where only the off-diagonal elements representing the important spatial correlations are parameterised and optimised during the variational inversion process. These include correlations between properties at different spatial locations, or between different physical parameters, that are known to interact strongly. All other off-diagonal elements, which represent weaker or less relevant correlations, are set to zero. This approach imposes a specific structure on $\mathbf{L}$, allowing the non-zero off-diagonal elements to be freely updated during optimisation. As a result, it yields a more efficient and tractable approach that still captures the important spatial correlations.

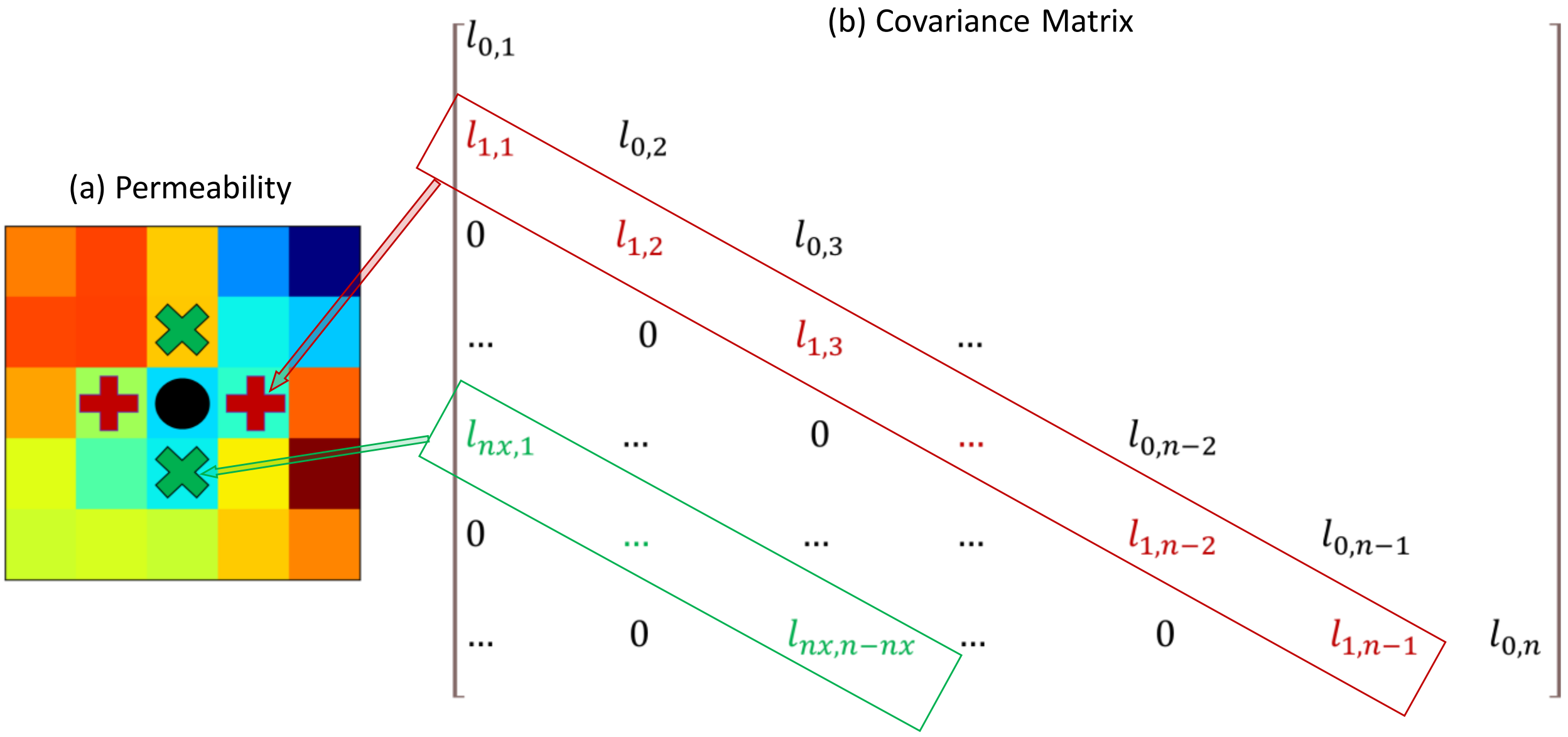

**Figure 2.** Construction of the physically structured covariance matrix. Red box represents correlations between horizontally adjacent permeability parameters (e.g., between the black circle and red "+" in the left panel), and green box shows correlations between vertically adjacent cells (e.g., between the black circle and green crosses).

The construction of the physically structured covariance matrix used here is illustrated in Figure 2. Suppose each cell (one is denoted by a black circle) is correlated with its four immediate neighbours. If we represent the 2D image with a 1D vector, and convert the image in a row-continuous manner, the red off-diagonal box in Figure 2b indicates correlations with horizontally adjacent cells (marked as red "+"), while the green box contains correlations with vertically adjacent cells (marked as green crosses). This structure can be expressed in the following equation:

$$
\mathbf{L} = \begin{bmatrix}
l_{0,1} & & & & & & \\
l_{1,1} & l_{0,2} & & & & & \\
0 & l_{1,2} & l_{0,3} & & & & \\
\dots & 0 & l_{1,3} & \dots & & & \\
l_{nx,1} & \dots & 0 & \dots & l_{0,n-2} & & \\
0 & \dots & \dots & \dots & l_{1,n-2} & l_{0,n-1} & \\
\dots & 0 & l_{nx,n-nx} & \dots & 0 & l_{1,n-1} & l_{0,n}
\end{bmatrix} \tag{16}
$$

The first subscript identifies a line of off-diagonal elements positioned a certain number of matrix rows below the main diagonal. For example, a subscript of 1 refers to one row below the main diagonal. The second subscript specifies the position of an element within that off-diagonal array of elements; for instance, a subscript of 1 refers to the first element within that array.

The off-diagonal elements positioned one row below the main diagonal represent correlations between horizontally adjacent grid cells, meaning those located next to each other along a row. Off-diagonal elements positioned $n_x$ rows below the main diagonal capture correlations between vertically adjacent grid cells, assuming that the model has width $n_x$ cells horizontally. To model interactions between cells that are two rows apart in the permeability model, off-diagonal elements located $2n_x$ positions below the main diagonal can be included, similarly for greater distances between cells, and in three dimensional models an equivalent approach is taken. These structured off-diagonal terms are essential for capturing spatial continuity and the underlying physical interactions among model parameters.

This reduces the number of parameters to optimise. For example, using a 5×5 kernel means each cell (the black circle) correlates with its 24 surrounding cells within a 5-by-5 square. In this case, the black circle correlates with all cells within the 5-by-5 permeability grid shown in Figure 2a. This results in one diagonal line plus up to 12 off-diagonal lines in the correlation matrix, totalling fewer than 13*n* non-zero elements, where *n* is the number of grid cells. This is much more efficient than full-rank ADVI, which requires that *n*(*n*+1)/2 parameters are optimised (Zhao & Curtis, 2024b). Unless otherwise stated, in variational optimisations the variational mean is initialised to zero and the Cholesky factor $\mathbf{L}$ is initialised as an identity matrix.

## 2.4 Workflow

To demonstrate the practical applicability of the variational inference methods discussed above, we now turn our attention to subsurface fluid flow inverse problems, specifically in the context of $CO_2$ sequestration. We introduce a workflow that applies different inference methods to estimate spatially varying model parameters from both direct and indirect observations. Figure 3 shows an overview of our workflow, which mainly includes three steps.

Step 1: Define the spatially varying permeability as the uncertain model parameter, and assign it a prior probability density function (pdf) to represent the initial knowledge before inversion.

Step 2: Construct a forward model that predicts well pressure and the heterogeneous seismic velocity from the permeability (and associated porosity) models by simulating $CO_2$ saturation and fluid pressure using a fluid flow forward function, then applying a rock physics model that links $CO_2$ saturation to seismic velocity.

Step 3: Perform inversion by comparing the predicted seismic velocity and well pressure to observed data using various inversion methods, including mean-field ADVI, PSVI, full-rank ADVI, Stein variational gradient descent (SVGD), stochastic SVGD (sSVGD), and Metropolis-Hastings Markov chain Monte Carlo (MH-McMC). Specifically, we use a pre-defined variational Gaussian family to approximate the target posterior pdf in mean-field ADVI, PSVI and full-rank ADVI (Kucukelbir et al., 2017; Zhao & Curtis, 2024b). For SVGD (Liu & Wang, 2016), the target posterior pdf is approximated using 300 particles. In the two Monte Carlo methods, sSVGD (Gallego & Insua, 2018) and MH-MCMC (Hastings, 1970; Metropolis et al., 1953), we employ 20 chains with iterative updates within each chain to generate posterior samples.

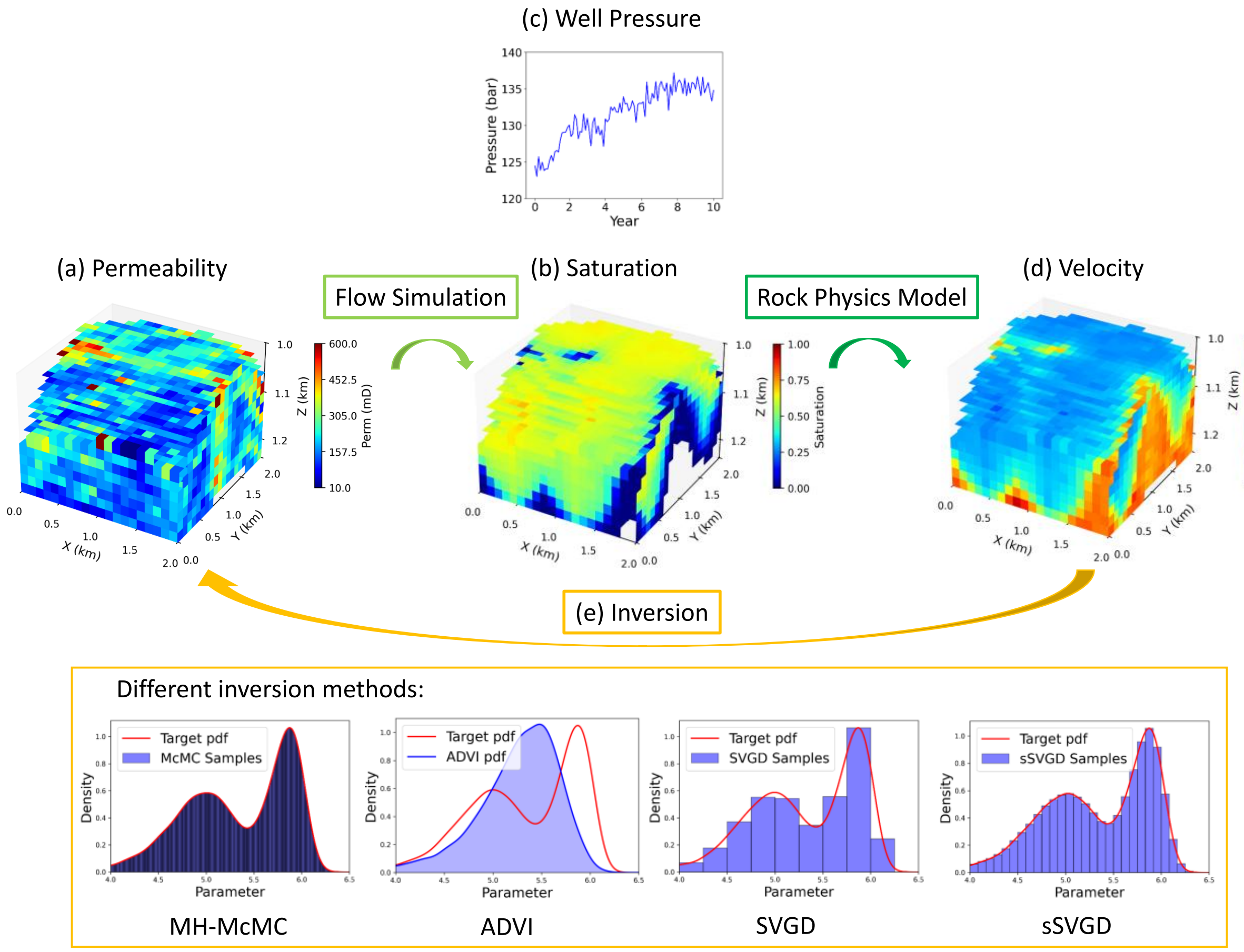

**Figure 3.** Workflow. Given a set of model parameters, specifically the permeability models in (a), the forward model simulates (b) fluid saturation and (c) well pressure. The predicted saturation is then input into a rock physics model to estimate (d) a seismic velocity model. In the inversion, these predicted seismic velocities and well pressures are compared with observed data to estimate the posterior distribution over permeability using (e) various inversion methods, including MH-McMC, ADVI, SVGD and sSVGD. Each method produces a different form of estimate: MH-McMC with infinitely many samples would produce a perfect rendering of the posterior distribution (but of course is far noisier than this in practice). ADVI produces an inverse log-transformed Gaussian distribution. In practice both SVGD and sSVGD produce an ensemble of model samples which may be binned and histogrammed to approximate the posterior pdf. SVGD produces far fewer samples than sSVGD, so finer bins can be used for the latter without compromising the accuracy of estimated probabilities.

## 3 $CO_2$ Sequestration Examples

### 3.1 Two-dimensional example

We begin by evaluating variational inference methods using a 2D $CO_2$ sequestration example which is small enough to allow methods to provide robust estimates of at least posterior mean structures. Although recovering exact posterior statistics over model parameters is infeasible, obtaining a reliable benchmark approximation is important in order for us to understand methodological comparisons. For this purpose, we use Metropolis-Hastings Markov Chain Monte Carlo (MH-McMC) with as large a number of samples as we could reasonably compute, a method which is known to yield asymptotically reliable posterior estimates as the number of iterations increases. For this small-scale 2D test, we believe that we generated a sufficiently large sample set to serve as an approximate reference.

The 2D reservoir model comprises a grid of cells with dimensions [20, 20], with each cell measuring 40 m and 20 m in the x and z directions, respectively. This setup results in an inverse problem involving 400 unknown model parameters. The true (isotropic) permeability $\mathbf{k}$ is generated using a Gaussian random field so as to embody suitable heterogeneity of subsurface structures at this scale (Müller et al., 2022). Model parameters $m$ are then defined as $m_i = k(\mathbf{x}_i)$, where each $\mathbf{x}_i$ takes the value of one of the locations of the grid of cells (in this case, $i = 1, \dots, 400$).

Figure 4a shows the true permeability models used to generate the observed data. The central purple line represents the injector, and the two red lines denote the two monitoring wells. Figure 4b highlights the points, log profiles and rectangular region referred to below. Figure 4c shows the geological correlation structure used to generate the true permeability field shown in Figure 4a.

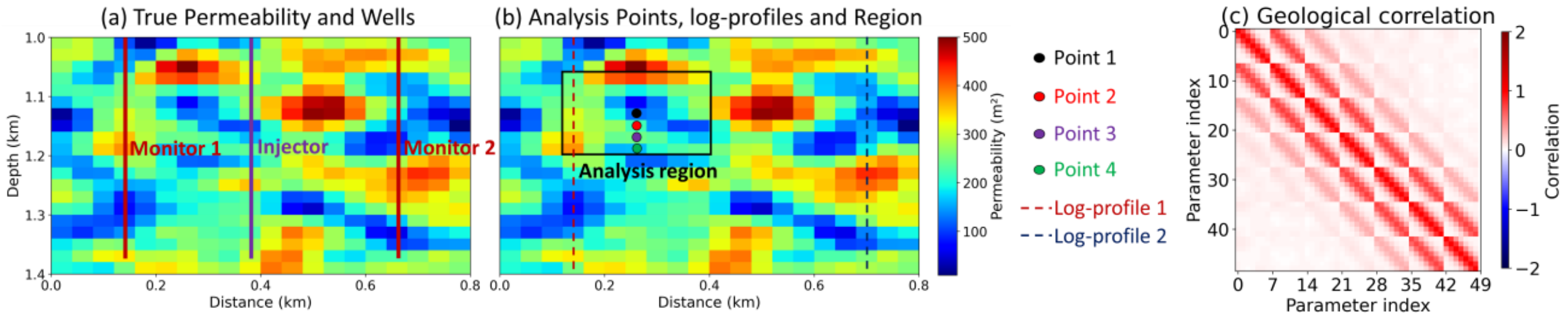

**Figure 4.** (a) The true permeability used to generate the observed data. The central purple line represents the injector, while the two red lines indicate the monitoring wells. (b) The analysis region (black rectangle) used to compute the correlation matrix in Figure 6, along with four points at which we analyse correlations between permeability parameters in Figures 7 to 9. The dashed lines represent the two log-profiles represented in Figure 10. (c) The geological correlation matrix used to generate the true permeability field.

A uniform prior pdf between 10 and 500 milli-Darcy (mD) is assigned to the permeability of each grid cell. The heterogeneous porosity $\boldsymbol{\phi}$ is derived from a permeability-porosity relationship calibrated from the Endurance field (BEIS, 2021):

$$\mathbf{k} = 10^{(4.19211 \times \log_{10} \boldsymbol{\phi} + 5.168417)} \tag{17}$$

The reservoir's top depth is set to 1,000 m, representing typical conditions for $CO_2$ injection into a deep saline aquifer where maintaining $CO_2$ in its supercritical state is essential and typically requires pressure and temperature conditions that occur at depths >800m (Bruant et al., 2002). The density and viscosity of $CO_2$ are set to be constant at 700 kg/m³ and 0.1 centipoise (cP), respectively, while the brine is assigned a density of 1,190 kg/m³ and a viscosity of 1 cP, consistent with fluid properties reported at a depth of 1,000 m for the Endurance field (BEIS, 2021). Endurance is a potential $CO_2$ storage site in the UK's Southern North Sea, which will be discussed in more detail in the 3D case study below. Injection is performed at the bottom of the single well in Figure 4a at a constant rate, which for this small example is 0.1 million tons per year over a five-year period. No-flow conditions are applied to the upper and lower boundaries to represent the sealing effect of surrounding rock formations, while constant-pressure conditions are imposed on the left and right boundaries to simulate open reservoir boundaries, allowing $CO_2$ to migrate laterally.

Pressure changes in the subsurface are monitored at a single point in the injector well at (190m, 1370m), and in two monitor wells at (70m, 1370m) and (330m, 1370m) in Figure 4a. One seismic velocity map is extracted annually, assumed to be derived from seismic inversion such as travel-time tomography or full waveform inversion including reliable Bayesian uncertainty estimates (Mosser et al., 2020; Zhang Xin et al., 2023; Zhao et al., 2022). Together, the pressure and seismic data form the observations used in the inversion. The data likelihood is assumed to be a Gaussian distribution, where the measurement uncertainties have standard deviations equal to 3% of the recorded pressure and 1% of the recorded seismic velocities. We add Gaussian noise from these same distributions to the data.

The fluid flow and data simulation forward problem is solved using the finite difference method (Moyner, 2024), coupled with a rock physics model that relates $CO_2$ saturation to seismic velocity once per year (Dhananjay Kumar, 2006). The governing equations for two-phase flow and the rock physics model are provided in Appendix B. To compute gradients of the objective function with respect to the model parameters efficiently, we use PyTorch to backpropagate gradients from the objective function through the data misfit, and then propagate the gradients to the model parameters via the adjoint-state method used in fluid flow simulations (Moyner, 2024; Paszke et al., 2019).

We perform inversion to estimate the posterior distribution of the permeability models using mean-field ADVI, PSVI, full-rank ADVI, SVGD, sSVGD and MH-McMC. While mean-field ADVI assumes independence among permeability parameters by using a diagonal covariance matrix, PSVI incorporates local spatial correlations by constructing the covariance matrix with a 5×5 spatial correlation kernel (Figure 2), inferring correlation coefficients between each permeability cell and its 24 surrounding neighbours. Full-rank ADVI captures correlations across the entire permeability models by optimising a full-rank covariance matrix.

For both ADVI and PSVI, we use the same setup to approximate the $\mathrm{ELBO}[q(\mathbf{m})]$ and its gradient: two Monte Carlo samples per iteration over a total of 4,000 iterations. For the SVGD methods, 300 particles are updated over 500 iterations to approximate the target posterior distribution. The sSVGD method is a stochastic McMC method which employs 20 chains, each running for 6,000 iterations. As a benchmark, we use a Metropolis-Hastings-based McMC algorithm with 20 chains, each running for 500,000 iterations to obtain a reliable posterior pdf approximation against which we can evaluate the performance of the variational inference methods, where reliability was checked by ensuring that independent random subsets of chains provided similar results to those shown below.

Figure 5 compares the six algorithms in terms of the estimated mean permeability models, standard deviations, and relative errors defined as the ratio of the absolute difference between the estimated mean and the true permeability to the standard deviation. It is important to note that the posterior mean should not necessarily match the true permeability models since the mean is a statistic of the entire posterior probability distribution rather than an estimator of the true solution. Nevertheless, in this case the inferred mean permeability models exhibit reasonably similar spatial patterns across all methods and resemble the true permeability. This overall agreement suggests that all inversion methods effectively combine information from the observed data and the prior, yielding more informative posterior approximations than the prior distribution.

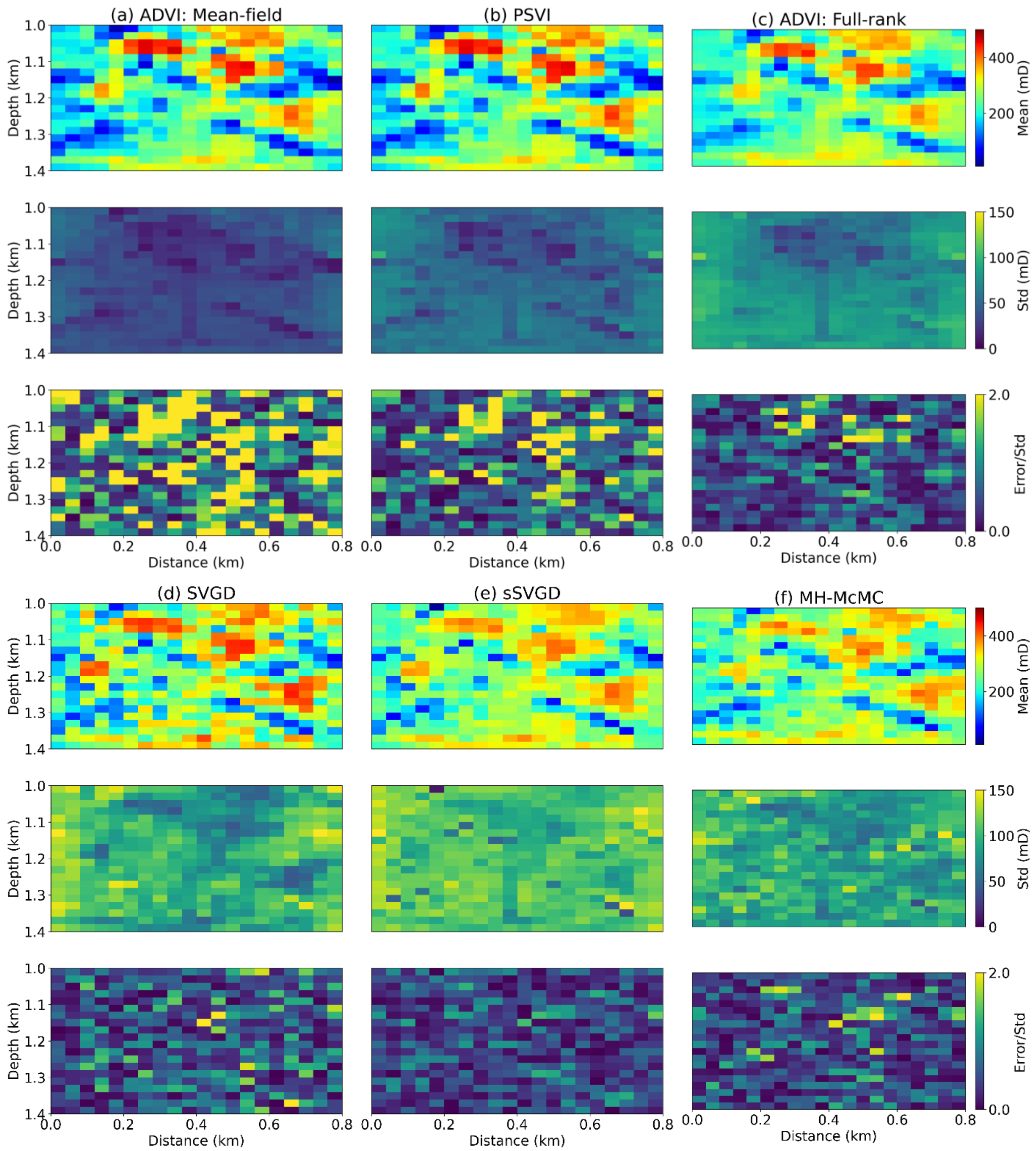

**Figure 5.** The mean (first row), standard deviation (std, second row), and relative error (third row), computed using (a) mean-field ADVI, (b) PSVI, (c) full-rank ADVI, (d) SVGD, (e) sSVGD and (f) MH-McMC. The relative error is the absolute value of the mean minus the true permeability, divided by the standard deviation.

Markov chain Monte Carlo (McMC) method provides asymptotically accurate approximations of the target posterior pdf given a sufficiently large number of samples. In this study, we address an inverse problem with 400 parameters, running 500,000 iterations per chain across 20 chains for MH-McMC. Given the extensive sampling effort and the fact that we observed stability of the estimated permeability mean and standard deviation after 500,000 iterations, we consider the MH-McMC results to be adequately converged to provide benchmark statistics for evaluating the performance of the other five methods. The mean permeability models inferred by all methods closely match the MH-McMC estimates, demonstrating their accuracy.

The second row shows the standard deviation maps of the permeability models, computed from samples drawn from the posterior pdf approximations produced by each of the six methods. Using the standard deviations estimated by the MH-McMC as a benchmark, we observe that the standard deviations obtained from the mean-field ADVI are significantly smaller. This does not imply that mean-field ADVI reduces parameter uncertainties more effectively; rather, the reduced uncertainties arise from the assumptions inherent to mean-field ADVI. Specifically, by assuming independence among permeability parameters by employing a diagonal covariance matrix, an unrealistically low estimate of uncertainties in the marginal distributions for each parameter are both expected and observed (Zhang Xin et al., 2023).

By incorporating correlations between each permeability cell and its 24 neighbouring cells using a 5×5 correlation kernel to construct the covariance matrix, PSVI yields substantially larger standard deviations than mean-field ADVI. This observation is further supported by the third row, where the relative error from PSVI is significantly smaller than that of mean-field ADVI, though slightly higher than those of the other four methods. Furthermore, its geometrical patterns of standard deviation closely resembles those in panels (c) from the full-rank ADVI method which accounts for correlations between each permeability cell and all other cells, which in turn produced reasonably similar patterns (with slightly smaller values) to the MH-McMC benchmark. These results suggest that constructing a sparse covariance matrix that captures dominant local correlations enables uncertainty estimation that is both more computationally efficient than full-rank ADVI, more realistic than assuming parameter independence, and produces patterns that match those of the true uncertainties.

To compare posterior correlations evaluated using different methods, we compute the posterior correlation matrix for a region highlighted by a black rectangle in Figure 4b, using posterior samples from the six methods (for the three ADVI-based methods we sample the resulting transformed Gaussian solution). The results are shown in Figure 6.

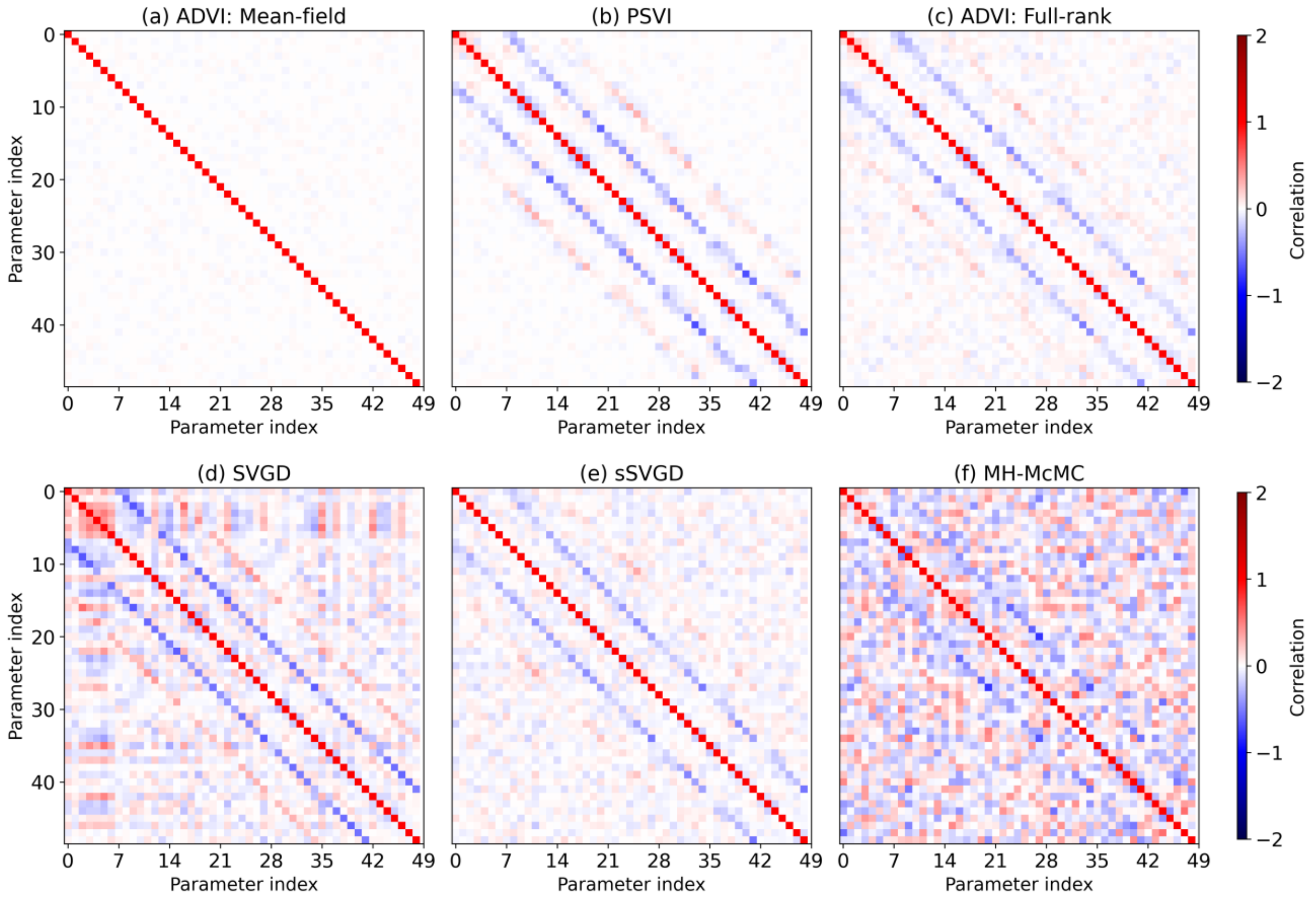

**Figure 6.** Correlation matrices for the analysis region inside of the black box in Figure 4b, computed using: (a) mean-field ADVI, (b) PSVI, (c) full-rank ADVI, (d) SVGD, (e) sSVGD, and (f) MH-McMC.

Figure 6 shows a variety of correlation matrices. It may be tempting to compare these to the correlation matrix shown in Figure 4c, however the two figures display fundamentally different information. The correlations in Figure 4c express the random distribution that was used to generate the single 'true' structure in our tests. In other words, it describes the variability in the set of geological worlds that might have existed in our cross-section, given geological processes operating over millions of years, and from which we select only one example to represent the true geology.

We note that if this correlation structure was known or estimated a priori, it could have been included as prior information in our Bayesian inferences (term $p(\mathbf{m})$ in equation 1). However, herein we have assumed that it is not known so it plays no further role in our solutions.

The correlations in Figure 6 are quite different: they describe residual model parameter uncertainty due to our limited ability to estimate the true permeabilities given only hard bounds on the velocities (uniformly distributed prior information) and the observed seismic and pressure data. The correlation structure in Figure 4 does not enter the solution, nor are the variational methods supposed to estimate it.

Specifically, as shown in Figure 6a, the correlation matrix computed from the posterior samples generated from the mean-field ADVI solution is almost diagonal. Theoretically it should be a diagonal matrix, but due to the limited number of samples used to estimate the correlation matrix, erroneous small off-diagonal values appear in the panel.

Figure 6b shows the correlation matrix derived from posterior samples generated by the PSVI using a 5×5 kernel to capture spatial correlations between permeability parameters. Distinct non-zero off-diagonal patterns appear. We analyse a 7×7 permeability grid as shown in Figure 4b, so the blue lines located 7 places from the diagonal line indicate negative correlations between each cell and cells in adjacent rows directly above or below. We also observe weaker blue lines located 1 place from the diagonal line, indicating weaker negative correlations between horizontally adjacent cells. Additionally, red off-diagonal lines appear 14 places from the diagonal, reflecting positive correlations between each permeability parameter and those two cells above or below it. This pattern of negative and positive correlations is identical in character to structures found in other spatial problems solved using seismic travel time tomography (Zhang & Curtis, 2020a) and full waveform inversion (Zhao & Curtis, 2024a).

Figures 6c and 6e show the correlation matrices computed from posterior samples generated by the full-rank ADVI and sSVGD methods, respectively. Both exhibit correlation structures similar to that obtained using PSVI with a 5×5 correlation kernel, suggesting that PSVI effectively captures the dominant correlations while maintaining a sparse representation of the covariance matrix.

Figure 6d shows the correlation matrix computed from SVGD and reveals a second blue off-diagonal line vaguely visible 21 places from the diagonal, indicating a potential negative correlation between permeability parameters separated by three rows and even a hint of correlation four rows apart. While it is unclear from this example alone whether this feature represents a true physical relationship or an artifact (it is not observed in the Monte Carlo solution in panel f), below we will show that in the 3D case a similar result may be related to the limited number of samples used in SVGD to approximate the high-dimensional posterior distribution.

Figure 6f shows the correlation matrix computed from MH-McMC. The negative correlations between vertically adjacent permeability parameters are clear, indicating negative vertical trade-offs. Interestingly, at the same off-diagonal locations, 14 places from the diagonal, where positive correlations appear in the PSVI, full-rank ADVI, SVGD, and sSVGD results, the MH-McMC matrix exhibits less obvious correlations. This is likely because those positive correlations are embedded within a background of positive and negative speckle noise observable throughout Figure 6f, making them harder to detect visually. The speckling varies when independent subsets of Monte Carlo samples are used, showing that while its mean, variances and the off-diagonal pattern 7 places from the diagonal are all relatively stable and reliable, other off-diagonal Monte Carlo correlation estimates are noisy due to convergence-related limitations in our benchmark solution. While 400 is a low number of parameters for a fluid flow tomography problem, this is still an extremely high dimensional nonlinear inverse problem when we consider that computationally intractable properties due to the curse of dimensionality emerge after only around 10 dimensions (Curtis & Lomax, 2001). The MH-McMC method can therefore not be expected to provide a perfect reference solution.

The correlation pattern suggests that neighbouring permeability cells are correlated with each other. One possible explanation lies in the nature of the observed data. Since well pressure measurements are only available at sparse well locations, and pressure propagates between injector and monitor wells over a significant portion of the domain, the pressure component of the data should be regarded as a remote sensing problem – analogously to seismic tomography methods (Zhang & Curtis, 2020a; Zhao et al., 2022). As a result, the impact on the data of an increase in permeability in one cell along the flow path might be able to be compensated by a decrease in permeability in an adjacent cell, together with possible increases in *their* neighbouring cells and so on; such variations would result in almost zero change to the spatially-averaged effective permeability, and hence to the observed well pressure. This compensatory effect implies that locally to each cell, model parameter estimates afforded by well pressure measurements have limited spatial resolution. The clear correlations observed in the posterior correlation matrices, in which the off-diagonal bands predominantly display negative or positive values, reflect this.

While the seismic velocity data in each cell are calculated from the values of fluid saturation and rock porosity in that cell alone, those fluid properties depend in turn on the flow simulation throughout the model. Their measured values in each cell can therefore be reproduced using permeability values in *other* cells that conform to exactly the same compensatory trade-offs as described above. Thus, every datum in this inverse problem can be considered to sense (provide information about) the permeability averaged over a set of local, remote cells, resulting overall in a remote sensing problem with limited spatial resolution. This explains why the appearance of local posterior correlations are similar to those observed in other types of remote sensing problems that have limited spatial resolution.

A more detailed discussion can be found in Appendix C, where we compare correlation matrices obtained when we use either only pressure data, or only velocity data, as observations during the inversion. The correlation matrices derived solely from pressure data exhibit predominantly negative correlations, which is consistent with the sparse nature of the pressure observations. On the other hand, the correlation matrices generated from velocity data shows both positive and negative correlations, presumably arising from a more localised trade-off among permeability parameters due to the spatially distributed nature of the observed velocity data. The results in Figure 6 comprise a combination of the two.

We also observe that correlations between permeability parameters are predominantly vertical rather than horizontal, as shown in Figure 6: the off-diagonal values located one place away from the diagonal exhibit weaker negative correlations than those that appear seven positions away. This likely reflects the physical behaviour of $CO_2$ injection into brine: since $CO_2$ is less dense than brine, the injected $CO_2$ tends to segregate upwards. However, with a closed upper boundary the plume cannot continue rising indefinitely and instead spreads laterally, so the overall transport need not be dominantly vertical. The predominantly vertical permeability trade-offs observed in the inversion therefore likely reflect limited identifiability of vertical connectivity (for example, between stacked layers).

To better understand the trade-offs indicated by the correlation matrices, we select four permeability parameters in cells labelled point 1 through point 4 in Figure 4b, and analyse their mutual dependence by estimating two-dimensional marginal distributions from histograms of the samples used above. Figure 7 presents joint histograms between points 1 and 2 for the six inference methods. Since mean-field ADVI assumes independence between permeability parameters, we note that it is inherently incapable of capturing correlations between model parameters: as shown in Figure 7a, variations in permeability at point 1 therefore have no effect on the value at point 2.

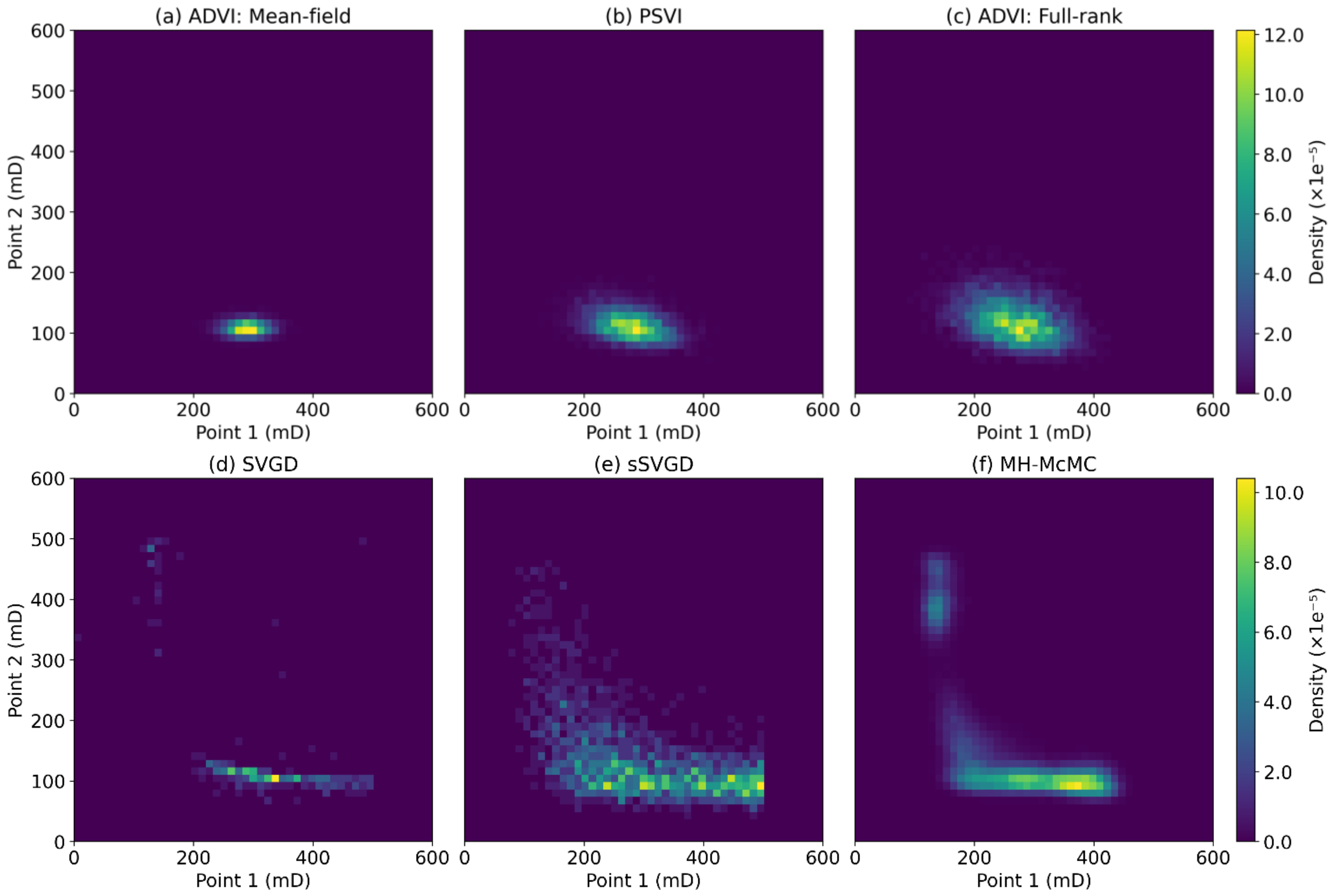

**Figure 7.** Joint marginal distributions (presented as normalised histograms) between permeabilities in vertically adjacent cells, points 1 and 2 in Figure 4b, based on posterior samples generated by (a) mean-field ADVI, (b) PSVI, (c) full-rank ADVI, (d) SVGD, (e) sSVGD, and (f) MH-McMC.

PSVI and full-rank ADVI exhibit similar correlation patterns in Figure 7b and 7c: as the permeability at point 1 increases, that at point 2 decreases, indicating a clear negative linear dependency. This is consistent with the blue off-diagonal lines in Figures 6b and 6c, which represent negative correlations between vertically adjacent permeability parameters. However, it is important to note that the Gaussian-shaped posterior pdf assumed in ADVI methods constrains statistical dependencies between parameters to be linear.

The statistical dependencies generated by SVGD, sSVGD, and MH-McMC are shown in panels (d), (e) and (f) respectively, in which an increase in permeability at Point 1 between 200 mD and 400 mD correlates with a slight decrease at Point 2. This negative correlation is again consistent with the correlation matrices presented in Figure 6, and is visually similar to those from PSVI and full-rank ADVI. However, for values below 200 mD at point 1, all three methods show a strongly nonlinear relationship between the two permeabilities, with a sharp increase in the value at point 2 to values around 400 mD (MH-McMC) or 500 mD (SVGD). In fact, this solution has very low probability (light blue colours) compared to the solutions with point 2 permeability around 100 mD (yellow colours), but nevertheless the consistency between these independent methods does indicate that there is some possibility of significantly higher values. Thus, we conclude that while PSVI and full-rank ADVI capture the main correlation between these parameters, and at least PSVI does so with significantly lower computational requirements compared to the methods in panels (d) to (f) as discussed below, this trades off with neglecting the small possibility that the nonlinear dependency is important.

Figure 8 shows the joint marginal distributions between permeabilities at points 1 and 3 which are offset by two rows of cells in Figure 4b. As indicated by the correlation matrices in Figure 6, this should correspond to the positive correlations between permeability parameters, represented by the red off-diagonal lines. As expected, mean-field ADVI assumes no correlation among model parameters, whereas both PSVI and full-rank ADVI (Figures 8b and 8c) display a weak positive correlation under a Gaussian posterior assumption.

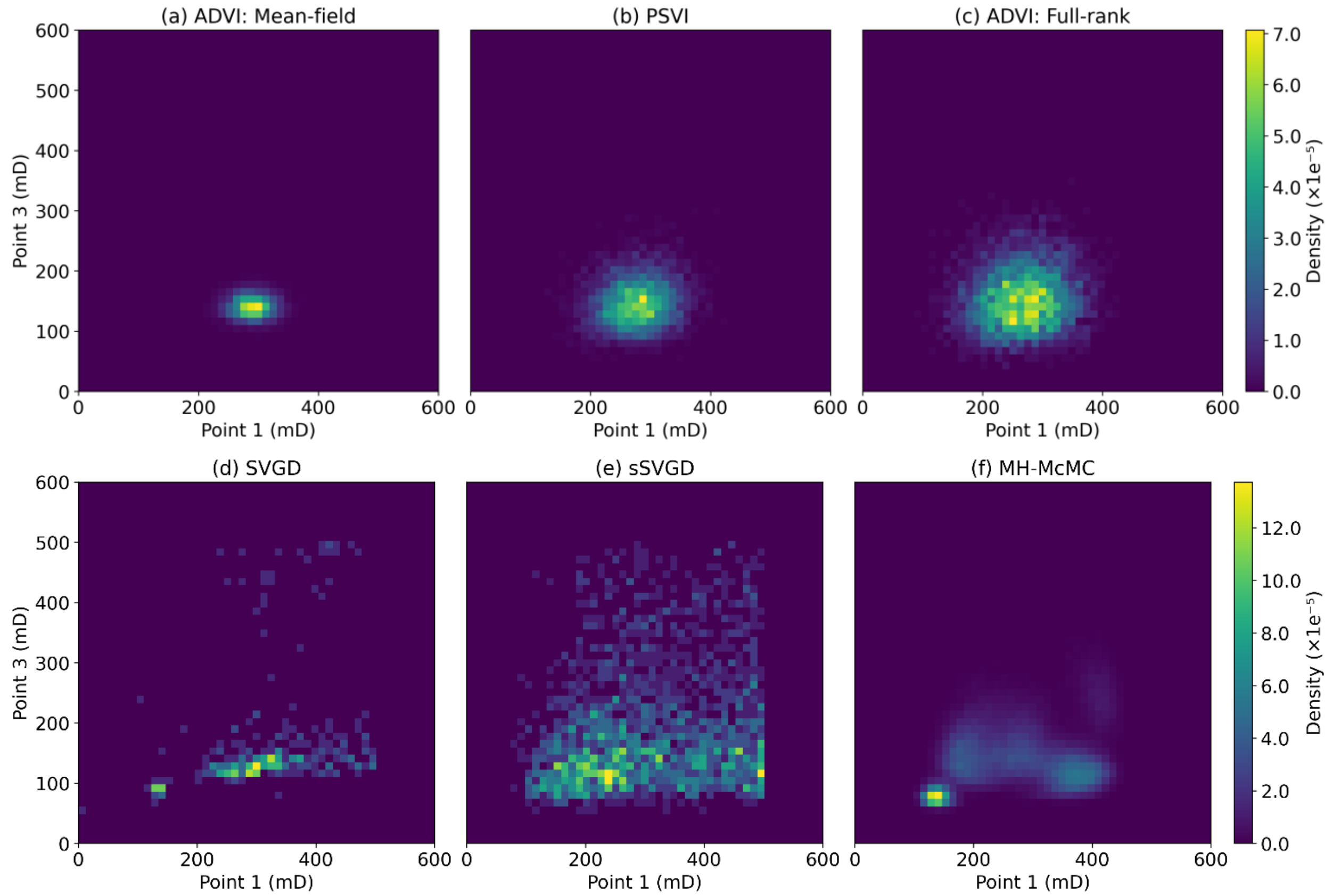

**Figure 8.** Similar to Figure 7, but now showing joint marginal distributions between permeabilities in vertically offset cells, point 1 and point 3 in Figure 4b.

The results from SVGD, sSVGD and MH-McMC in Figures 8d to 8f reveal a more scattered, and perhaps slightly non-linear positive correlation. The patterns produced by SVGD and sSVGD resemble those from MH-McMC to the extent that they all indicate a levelling off (and perhaps reversal) of the positive correlation beyond permeabilities of 300 mD at point 1, but their details do not match, and probably because of the lower number of samples (particles) produced by SVGD, this method shows far narrower uncertainties than any of the other methods. Overall, these results indicate again that while the ADVI methods return the principal features of the posterior marginal distribution, the other methods may capture additional details (at substantially increased cost).

Figure 9 shows joint marginal distributions between permeabilities at points 1 and 4, separated by three rows of cells in Figure 4b. As indicated by the correlation matrices in Figure 6, correlations at this distance should be negligible. Consistent with this, Figure 9 shows no clear correlation trend between point 1 and point 4 across all six inference methods. Notably, the posterior samples from the three ADVI methods are more tightly clustered than those from SVGD, sSVGD and MH-McMC, but there is substantial disagreement between results from the latter three methods so we suggest that at least those from SVGD and sSVGD cannot be trusted.

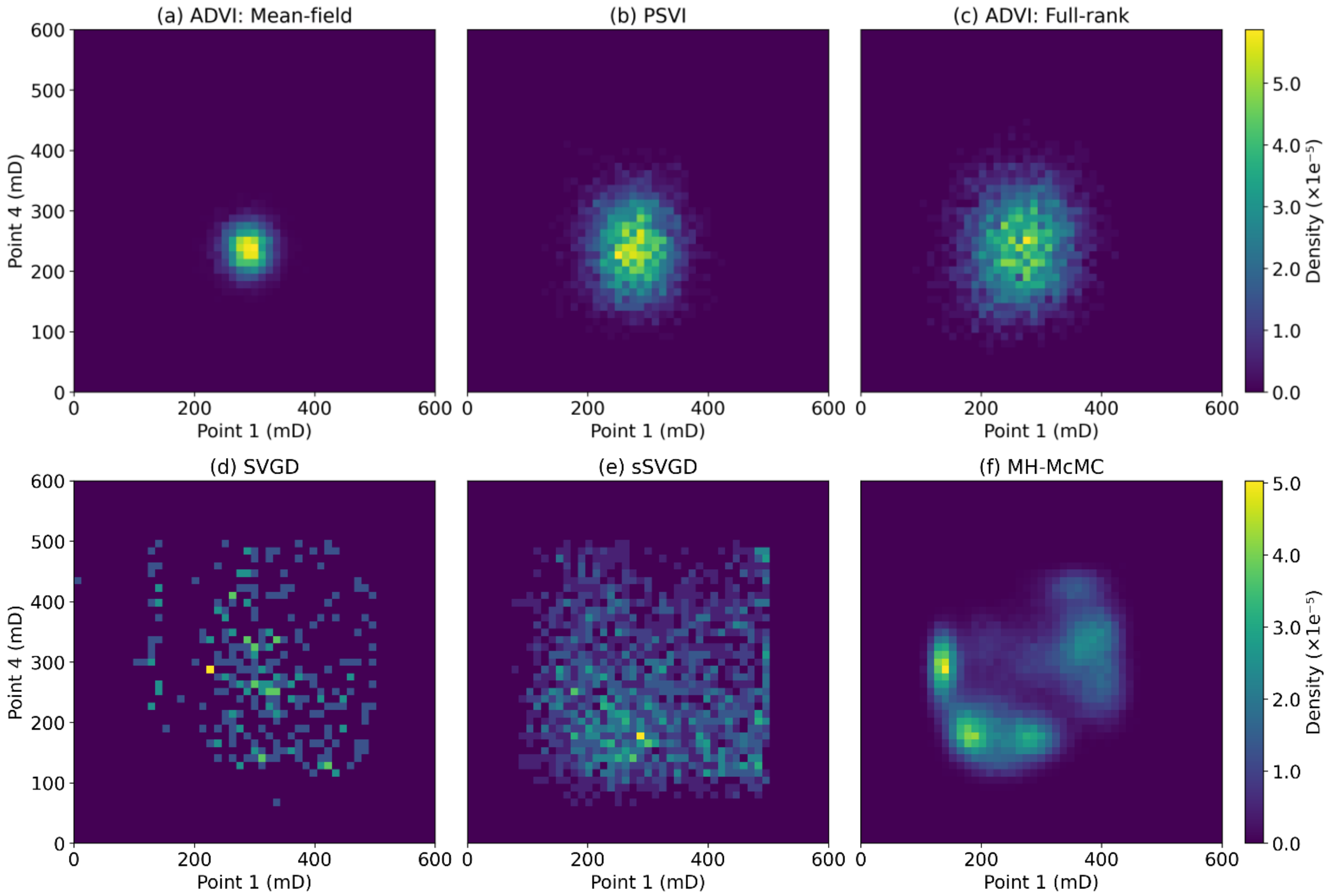

**Figure 9.** Similar to Figure 7, but now showing joint marginal distributions between permeabilities in vertically offset cells, point 1 and point 4 in Figure 4b.

To further compare details of the different posterior distributions obtained, Figure 10 presents two vertical profiles of the permeability models, starting from cells located at (140m, 1000m) and (700m, 1000m), as marked by the dashed lines in Figure 4b. Figure 10a shows that the posterior uncertainties along both profiles derived from mean-field ADVI are generally far narrower than that of the other five methods, reflecting mean-field ADVI's tendency to underestimate uncertainties due to its assumption of independence among permeability parameters. Notably, the true permeability values on profile 1 present very low posterior probability across depths of 1.1–1.14 km, and on profile 2 between 1.02–1.06 km, as indicated by the white arrows, again highlighting the method's limitations.

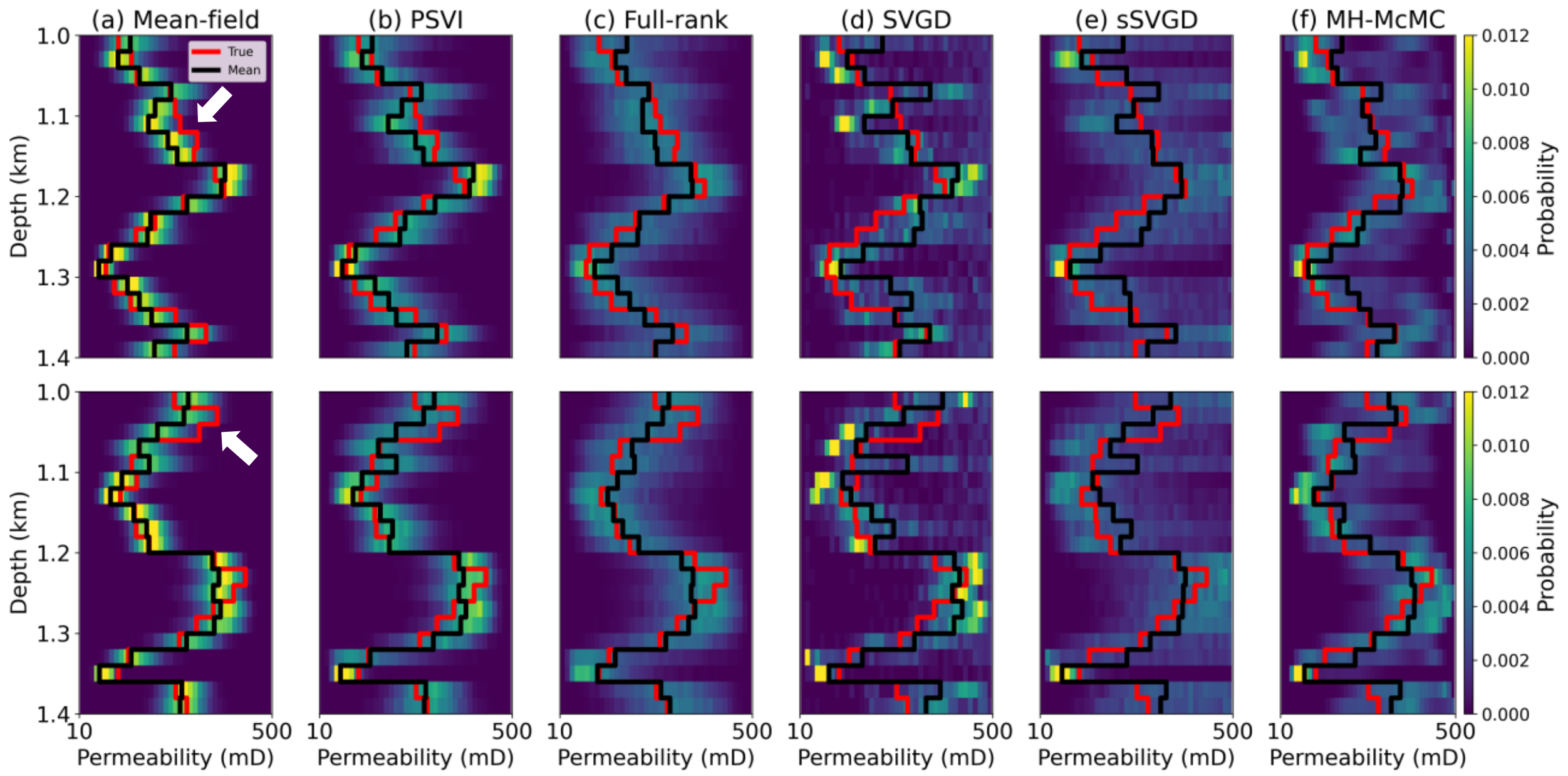

**Figure 10.** Marginal posterior distributions over permeabilities along vertical profiles 1 (top) and 2 (bottom) shown in Figure 4b, produced by different inference methods (column titles). Yellow and blue regions indicate high and low posterior probability, respectively. The red line shows the true permeability, and the black line shows the posterior mean permeability. The white arrows highlight regions where the true permeability lies in areas of low posterior probability.

By accounting for correlations only among spatially nearby permeability parameters, PSVI provides improved uncertainty estimates compared to mean-field ADVI, with all true permeability values falling within the range of significantly non-zero probabilities as observed on this colour scale (Figure 10b), similarly to full-rank ADVI (Figure 10c).

We observe that the posterior uncertainties estimated by SVGD and sSVGD in Figures 10d and 10e closely resemble each other, and are fairly similar to those in Figure 10f. They appear somewhat rougher than those from the three ADVI methods, presumably in part due to their reliance on still sparse samples to produce these sample-density plots, but since SVGD, sSVGD and MH-McMC are in principle capable of capturing multimodal, complex posteriors, it is possible that some of this roughness is correct.

Overall, results from the 2D tests indicate as expected that the ADVI methods produce less detailed estimates of the target posterior pdf due to their Gaussian assumptions. However, they offer significantly greater computational efficiency compared to SVGD, sSVGD and MH-McMC. Table 1 summarises the computational expense required by each method. Wall-clock time was measured on Intel(R) Xeon(R) Silver 4314 CPUs @ 2.40 GHz.

Table 1. The computational resources used in each inversion for the 2D test.

| Inversion methods | Number of simulations | Wall-clock time | Number of CPUs |
|---|---|---|---|
| Mean-field ADVI | 16,000 | 1.76 hours | 2 |
| PSVI | 16,000 | 2.78 hours | 2 |
| Full-rank ADVI | 16,000 | 3.28 hours | 2 |
| sSVGD | 240,000 | 11.86 hours | 20 |
| SVGD | 300,000 | 14.72 hours | 20 |
| MH-McMC | 10,000,000 | 833.34 hours | 20 |

Each method was run several times, and the results reported here represent typical numbers of simulations, wall-clock time, and the number of CPU's required to obtain a reasonably stable solution. The three ADVI variants (mean-field ADVI, PSVI and full-rank ADVI) were run on two CPUs because two Monte Carlo samples were used to estimate the ELBO. By contrast, sSVGD, SVGD and MCMC were run on twenty CPUs, reflecting the maximum computational resources available to us. Greater parallelisation would be expected to reduce wall-clock time further, but the values reported here provide an initial assessment under the current experimental set-up.

We therefore compare computational efficiency across methods using the number of simulations, thereby reducing the influence of hardware constraints. All variational inference methods required an adjoint run to estimate gradients of the forward model with respect to parameters, which has approximately the same computational cost as a forward run, whereas MCMC does not. To enable a fair comparison, we therefore count each variational inference run as two simulations (one forward and one adjoint) and compare this total with the number of forward runs in MCMC.

Specifically, the mean-field ADVI, PSVI and full-rank ADVI methods were each run for 4,000 iterations using 2 Monte Carlo samples per iteration, resulting in a total of 8,000 forward simulations and 8,000 adjoint simulations per method. The sSVGD algorithm used 20 chains, each optimised over 6,000 iterations, yielding 120,000 forward simulations and 120,000 adjoint simulations. SVGD employed 300 particles over 500 iterations, totalling 150,000 forward simulations and 150,000 adjoint simulations. For MH-McMC, 20 chains were run with 500,000 forward simulations each, leading to a total of 10,000,000 forward simulations. The ADVI methods exhibit significantly higher computational efficiency, requiring only 6.6%, 5.3% and 0.16% of the forward runs used by sSVGD, SVGD and MH-McMC, respectively. This efficiency is especially important in applications such as $CO_2$ sequestration, where each forward simulation is computationally expensive because it involves solving complex fluid flow equations.

The above results required a subjective evaluation of the number of samples required to reach a stable solution. To remove potential subjective biases, and further demonstrate the computational advantage of the ADVI methods, we also present inversion results obtained using the same number (8000) of forward simulations for all methods in Appendix E.

PSVI and full-rank ADVI allow non-zero correlations among model parameters, producing posterior covariance structures that better reflect the spatial dependencies expected in this problem and that agree more closely with the McMC reference results than the diagonal covariance assumed in mean-field ADVI. Any preference for PSVI over full-rank ADVI may seem unjustified since both require the same number of forward simulations (8,000), and full-rank ADVI captures the complete correlation structure. However, PSVI optimises 25,600 variational parameters whereas full-rank ADVI optimises 80,600. While this difference is not necessarily significant in this small, spatially-2D problem, in the spatially-3D case below we demonstrate that the number of variational parameters presents a major computational challenge for the full-rank method. In principle SVGD, sSVGD and MH-McMC produce more accurate estimates of posterior pdf, but this is only true at substantially increased cost compared to PSVI and full-rank ADVI. When the same number of samples are afforded each method, these three methods perform relatively poorly (Figure E1 and E2). Thus, our provisional conclusion is that PSVI stands out in our results: compared to the other methods, it offers lower optimisation complexity and greater computational efficiency, while still providing reasonable estimates of global parameter correlations and being extendable to much larger scale problems.

### 3.2 Three-dimensional example

We next evaluate the performance of the variational inference methods in a 3D $CO_2$ sequestration scenario constructed using published information about the Endurance $CO_2$ store – an anticline within the Bunter Formation in the UK's Southern North Sea. A $CO_2$ sequestration project at this site is scheduled to commence in 2026, with Phase 1 involving four injection wells, each operating at a rate of 1 million tonnes of $CO_2$ per year (BEIS, 2021). A reservoir model for the Endurance field was developed by Wenck et al. (2025) based on data from the Endurance report (BEIS, 2021). They generated the reference ('true') permeability field indirectly by first constructing a spatially correlated porosity field using ellipsoidal averaging over a moving 2D elliptical neighbourhood (Durlofsky et al., 1997) following Jackson & Krevor (2020). The porosity variance is calibrated by fitting an extreme value distribution to well 42/25d-3 porosity data (BEIS, 2021). Spatial correlation is imposed via correlation lengths $(r_x, r_y, r_z) = (1000\text{ m}, 760\text{ m}, 4\text{ m})$, where $r_x$ and $r_y$ are taken from BEIS Endurance report and $r_z$ is estimated from a semi-variogram of the well data (BEIS, 2021). Permeability is then computed using the porosity-permeability trend calibrated from well data (equation 17). The simulation model was subsequently upscaled to capture the effects of capillary heterogeneity on multiphase flow behaviour (Wenck et al., 2025).

Simulating the full-scale $CO_2$ fluid flow model required several hours of compute wall-clock time on our machines, which would make this study intractable. To reduce computational cost, we first applied vertical upscaling using a weighted harmonic averaging method (Salazar, 2007), and then cropped the reservoir to a grid of size (20, 20, 12) with cell size (100 m, 100 m, 20 m), where the third cell index is depth. Inactive grid cells were assigned to represent the anticline structure, resulting in 3,693 active permeability parameters with permeability values ranging from 10 mD to 600 mD (Figure 11a). $CO_2$ is injected through the bottom of four vertical wells placed approximately at planned locations, specifically in grid cells (450m, 250m, 1070m), (550m, 1550m, 1070m), (1250m, 350m, 1070m) and (1350m, 1450m, 1070m), each operating at a constant injection rate of 1 million tonnes per year over a ten-year period. Seismic velocity maps are simulated every two years using the rock physics model introduced above, yielding five maps over the injection period. Pressure data are recorded at the injection wells. To account for measurement uncertainties, we add 3% Gaussian noise to the well pressure data and 1% Gaussian noise to the velocity data. The remaining model setup, such as the initial pressure, temperature and boundary conditions, follows the configuration used in the 2D case. Figure 11a shows the true permeability models used to generate the observed data, along with the locations of the four injection wells. Figure 11b shows the two log-profile locations and the analysis region used to compute the correlation matrices in subsequent analysis.

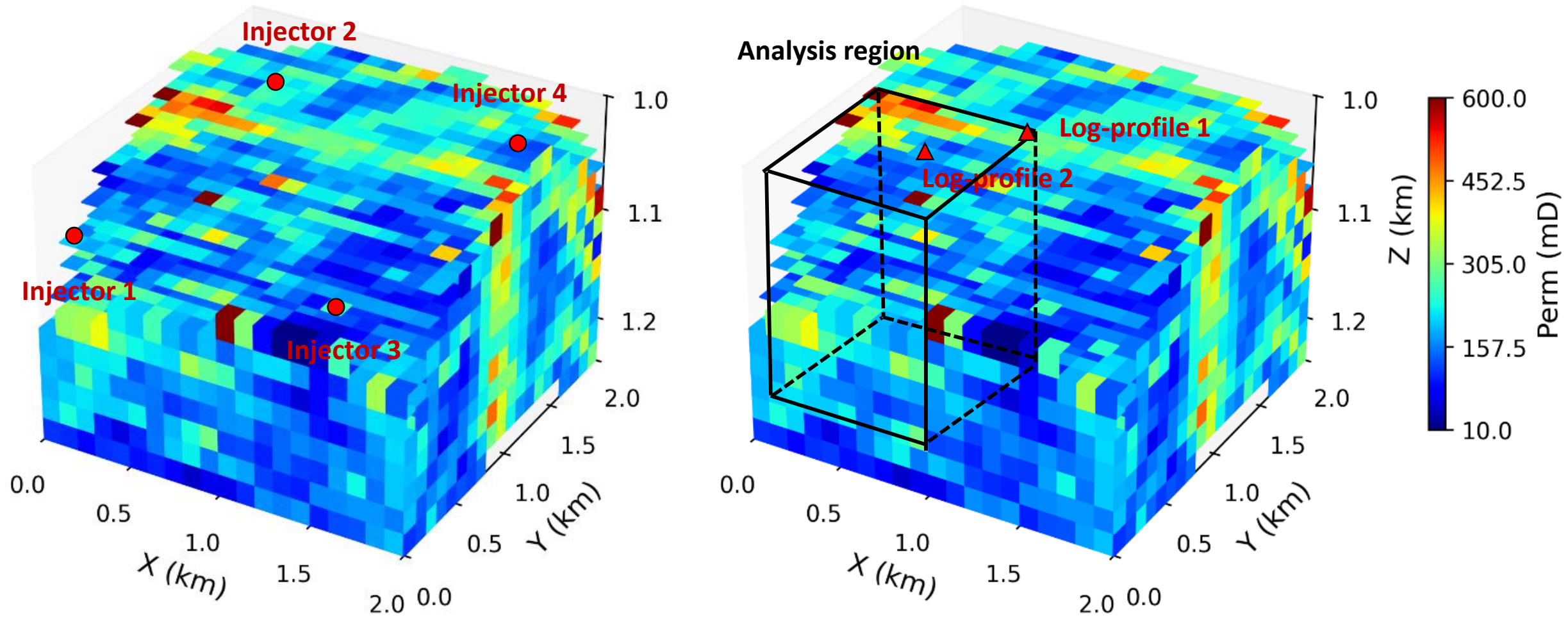

**Figure 11.** (a) The true permeability models used to generate the observed data, along with the locations of the injection wells. (b) The analysis region used to compute the correlation matrix (black rectangular box) shown in Figure 16. The two triangles indicate the locations of the log-profiles used in Figure 17.

Neither full-rank ADVI nor our Metropolis-Hastings McMC method could be run for sufficiently many iterations to converge to robust results in this high-dimensional problem (in repeated tests, substantially different solutions were found) due to limitations in available computing power and the curse of dimensionality. We therefore evaluate the performance of the other methods tested above, without having a reference solution in the 3D case. Our assumption is that if independent methods using different parametrisations produce similar posterior statistics, then our confidence in those statistics increases. The comparisons below should nevertheless be interpreted as relative assessments of consistency among different inference methods and physically plausible correlation structures, rather than definitive statements about absolute posterior accuracy.

Specifically, we apply mean-field ADVI, and PSVI with 5×5×5 (3D equivalents of the 5×5 kernel in Figure 2a), 7×7×7 and 9×9×9 correlation kernels. Each of these ADVI variants is optimised over 6,000 iterations using 2 Monte Carlo samples per iteration to find the optimal Gaussian variational distribution. For SVGD, we use 300 particles and perform 800 iterations to transform the particles toward the target posterior pdf iteratively. The sSVGD method employs 20 chains, each sampling over 8,000 iterations.

Figures 12a to 12f present the mean permeability estimates obtained using mean-field ADVI, PSVI with 5×5×5, 7×7×7 and 9×9×9 kernel sizes, SVGD and sSVGD, respectively. All six methods generally capture the main structural features of the reference field. However, discrepancies remain between the estimated and true permeability, which is expected given the limited and noisy observations and the non-uniqueness inherent in high-dimensional inverse problems.

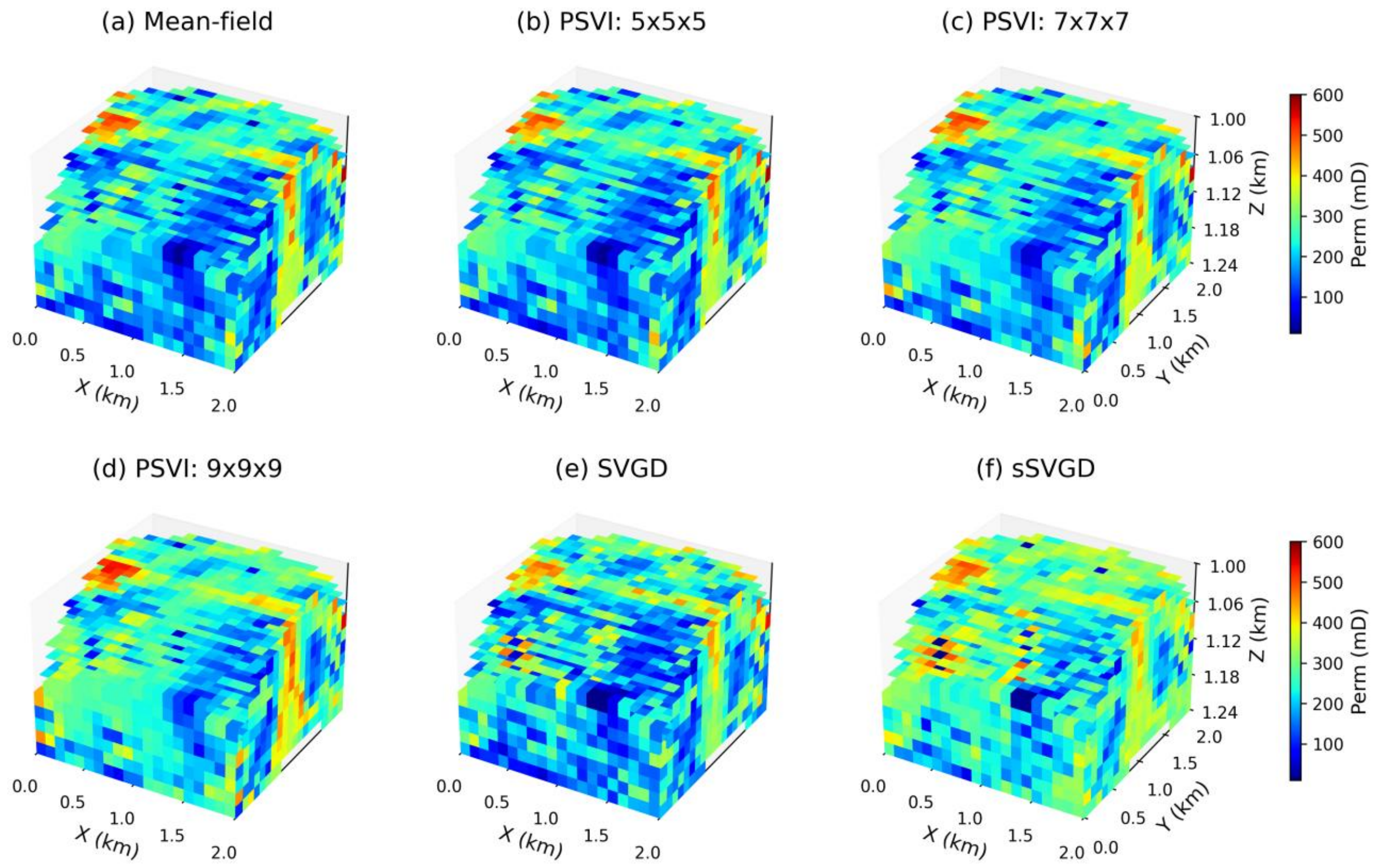

**Figure 12.** Mean permeability estimates using (a) Mean-field ADVI, (b) PSVI with a 5×5×5 kernel, (c) PSVI with a 7×7×7 kernel, (d) PSVI with a 9×9×9 kernel, (e) SVGD, and (f) sSVGD.

Figure 13 presents the standard deviation of the permeability models computed from the posterior samples generated by the same six methods. Figures 13a to 13d show that the standard deviations from the ADVI methods are generally smaller than those from the SVGD and sSVGD methods. Among the ADVI variants, the standard deviation increases slightly as we progress from mean-field ADVI to PSVI with increasing kernel sizes. Mean-field ADVI assumes independence among all model parameters which reduces posterior variability, while PSVI introduces structured covariances: the 5×5×5 kernel accounts for correlations between each cell and its 124 neighbours, the 7×7×7 kernel extends this to 342 neighbours, and the 9×9×9 kernel to 728 neighbours. As the kernel size increases, PSVI can represent a broader range of spatial correlations among model parameters, which can improve the representation of posterior dependencies and marginal uncertainties. However, this also means that more variational parameters need to be optimised during inversion, making the optimisation problems more difficult to solve.

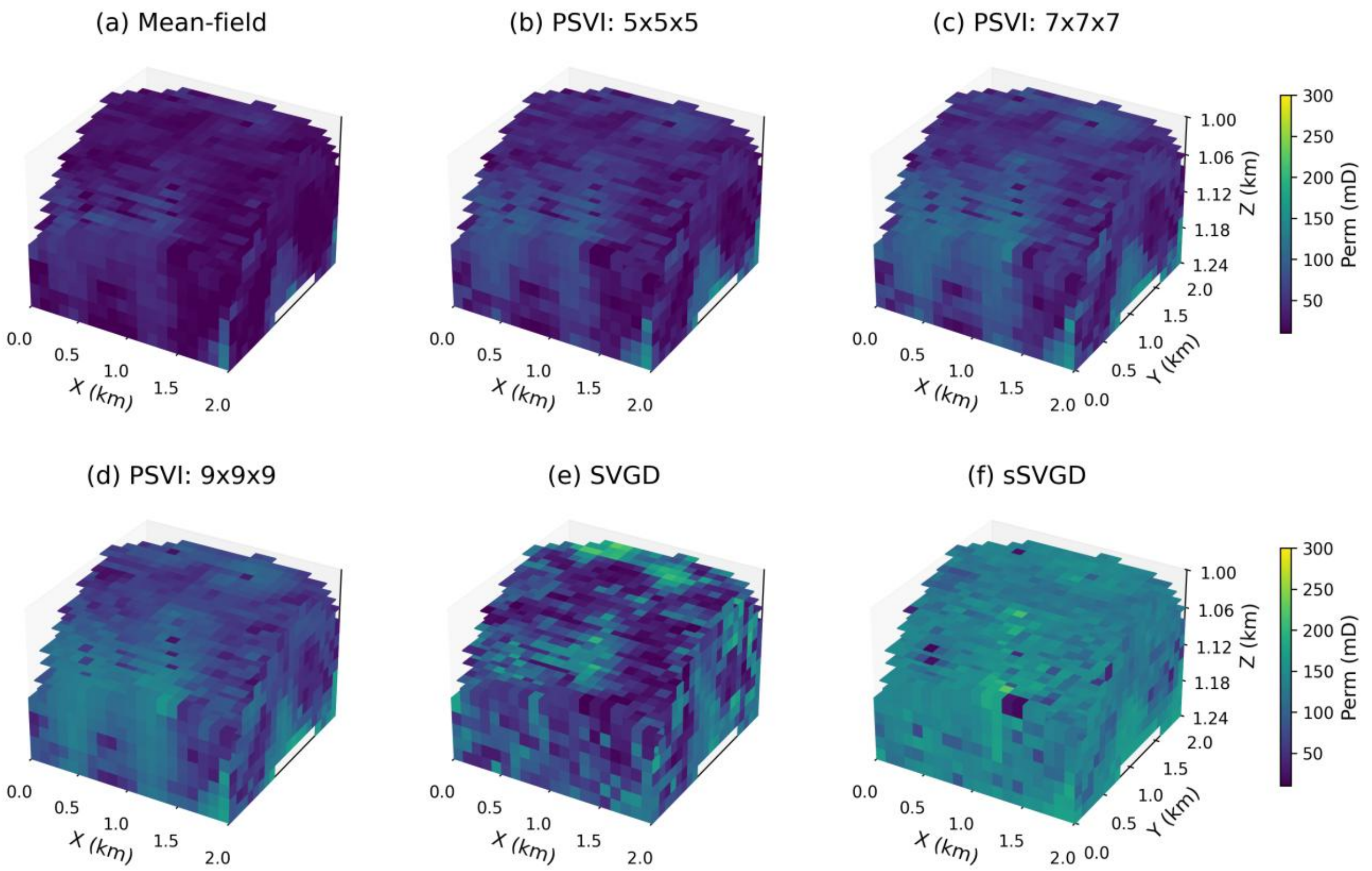

**Figure 13.** Standard deviation of the permeability models estimated from posterior samples using (a) mean-field ADVI, (b) PSVI with a 5×5×5 kernel, (c) PSVI with a 7×7×7 kernel, (d) PSVI with a 9×9×9 kernel, (e) SVGD, and (f) sSVGD.

Following this logic, where increasing the correlation kernel size between each cell and its neighbours leads to a more realistic posterior representation, full-rank ADVI should in principle provide the most representative covariance structure among the ADVI variants, as it captures all pairwise correlations across the parameter space. However, in practice, optimising full-rank ADVI with a full covariance matrix is extremely challenging: in this example the optimisation was unstable and failed to converge due to the large number of parameters required to model the full covariance structure (Zhao and Curtis 2025a, 2025c).

To demonstrate this, Figure 14a shows the full-rank ADVI results using 6,000 iterations and 2 Monte Carlo samples per iteration, the same settings applied to the mean-field ADVI and PSVI methods. The mean permeability estimated by full-rank ADVI deviates noticeably from the true permeability shown in Figure 11a and performs significantly worse than the mean permeability estimates from mean-field ADVI and PSVI shown in Figures 12a to 12d. This issue arises primarily because mean-field ADVI, PSVI with 5×5×5, 7×7×7 and 9×9×9 kernel sizes, and full-rank ADVI involve the estimation of 7386, 307200, 830400, 1756800 and 6824664 unknown variational parameters, respectively. The optimisation of full-rank ADVI is therefore substantially more computationally demanding than that of the mean-field ADVI and PSVI methods.

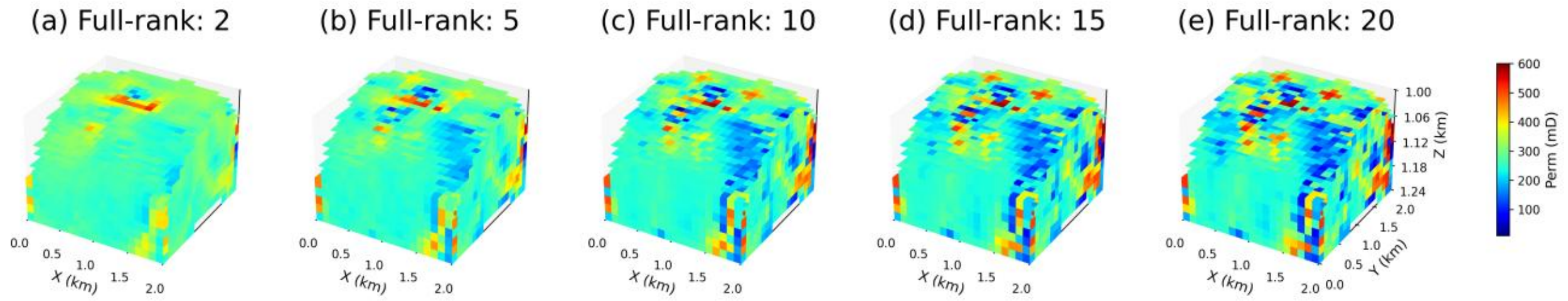

**Figure 14.** Estimated mean permeability models obtained using full-rank ADVI with different numbers of Monte Carlo samples per iteration: (a) 2 samples, (b) 5 samples, (c) 10 samples, (d) 15 samples, and (e) 20 samples.

To improve the convergence of full-rank ADVI, we can either increase the number of optimisation steps or increase the number of Monte Carlo samples per step, as a larger number of Monte Carlo samples theoretically provides more accurate gradient estimates. Since Monte Carlo sampling can be efficiently parallelized during forward simulation, we assess the convergence performance of full-rank ADVI using different numbers of Monte Carlo samples per iteration.

Figures 14b to 14e present the mean permeability estimates obtained using full-rank ADVI with 5, 10, 15 and 20 Monte Carlo samples per iteration over 6,000 iterations. As the number of Monte Carlo samples increases, the mean permeability estimates become more reliable, gradually converging slightly more towards the true permeability models. However, even with 20 Monte Carlo samples per iteration, the estimate still falls short of the similarity achieved by the other methods shown in Figure 12 and remains noticeably different from the reference field, so we assume that it is still substantially in error. This highlights the advantage of PSVI, which provides a more reliable posterior approximation by capturing important correlations compared to mean-field ADVI, while being far more tractable to optimise than full-rank ADVI.

The standard deviations from SVGD in Figure 13e are on average about the same size as those from the PSVI method with a kernel size of $9 \times 9 \times 9$ (Figure 13d), yet it is much lower than the standard deviation of sSVGD shown in Figure 13f. Neither SVGD nor sSVGD rely on the assumption of a Gaussian-shaped posterior pdf, so in principle they should produce relatively similar standard deviations.

To investigate this further we plot histograms of the permeability correlations between point 1 and points 2 to 4, where these are defined as follows. The first row in Figure 15 indicates that the permeability at point 1 (cell (6,6,6)) is negatively correlated with point 2, which corresponds to the permeability parameters in the cell below it. The second row shows a positive correlation between point 1 and point 3, representing permeability two cells below it. The third row shows no distinct correlation between point 1 and point 4, corresponding to three cells below it. These results are consistent with the correlation matrices shown below, and align with the correlation structure observed in the 2D case because the governing physical processes are essentially the same in both examples.

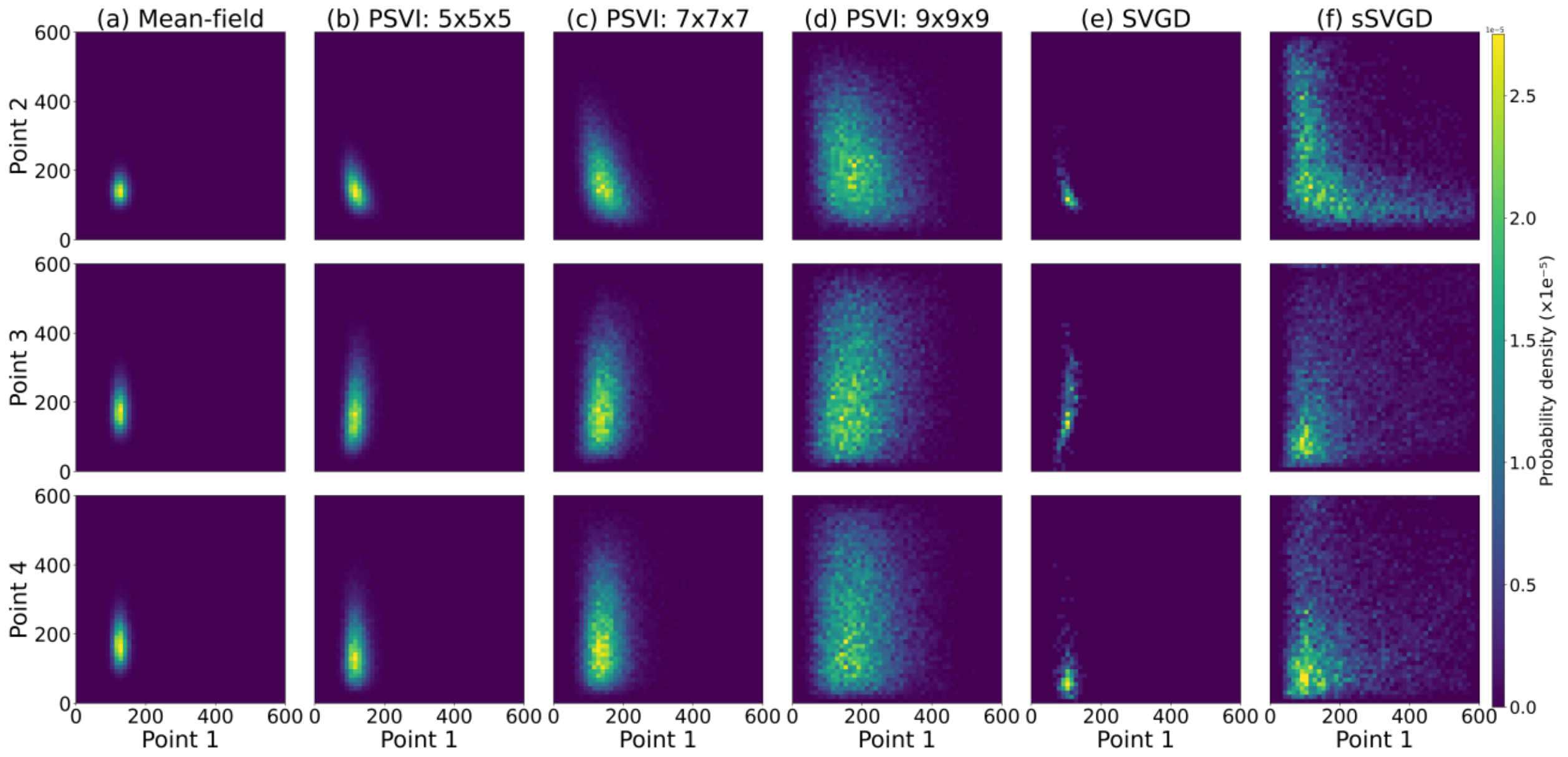

**Figure 15.** Joint marginal posterior distributions (plotted as normalised histograms) between the permeabilities in vertically offset cells located at point 1 (6, 6, 6), point 2 (6, 6, 7), point 3 (6, 6, 8), and point 4 (6, 6, 9): point 1 and point 2 (first row), point 1 and point 3 (second row), and point 1 and point 4 (third row), based on posterior samples generated by (a) mean-field ADVI, (b) PSVI with a 5×5×5 kernel, (c) PSVI with a 7×7×7 kernel, (d) PSVI with a 9×9×9 kernel, (e) SVGD, and (f) sSVGD.

The SVGD results in Figure 15e show posterior samples that are highly concentrated in a small region which is inconsistent with all results other than mean-field ADVI. Since the latter is known to produce uncertainties that are too low, this result reflects mode collapse in SVGD, in which all particles collapse into a narrow area and fail to adequately represent the full shape of the target posterior pdf (Angelo & Fortuin, 2021). When a limited number of particles is used in a high-dimensional inverse problem, the distances between particles are relatively large in high-dimensional space since almost all particles end up in different corners of the hyper-cubic parameter space (Curtis & Lomax, 2001). Within the SVGD algorithm there is a term which attracts particles to regions of high probability, and a mutually-repulsive component which should keep particles apart (Liu & Wang, 2016); in high dimensional cases where particles are always far apart, the attraction term dominates. This drives the particles toward regions of high posterior probability and soon causes them to collapse into an area where the likelihood is higher. They then fail to represent the full target posterior distribution, leading to underestimated standard deviation values as shown in Figure 15e. A detailed explanation of the mode collapse mechanism in SVGD is provided in Appendix D.

Figure 16 shows the correlation matrices for a subregion within the black box, spanning the permeability cell located at (100m, 100m, 1020m) to (1000m, 1000m, 1200m) as shown in Figure 11b. Specifically, Figure 16a shows the correlation matrix computed from posterior samples generated using the mean-field ADVI method: the diagonal structure reflects the mean-field assumption, where each diagonal element represents the variance of an individual model parameter. Figures 16b to 16d display the correlation matrices derived from posterior samples generated by PSVI with 5×5×5, 7×7×7 and 9×9×9 kernels, respectively. All three matrices exhibit a similar banded structure, with five distinct correlation bands visible. The central diagonal band represents the correlation between each permeability cell and its left and right neighbours. The two inner off-diagonal bands, centred around blue lines 81 places from the main diagonal, correspond to correlations with permeability parameters located one cell above and below. The outer off-diagonal bands, centred around red lines 162 places from the diagonal, reflect correlations with cells located two cells above and below. Similarly to the 2D case, the horizontal correlations appear to have smaller magnitudes than the vertical ones, likely due to the dominant vertical flow paths driven by the density contrast between the injected $CO_2$ and the brine.

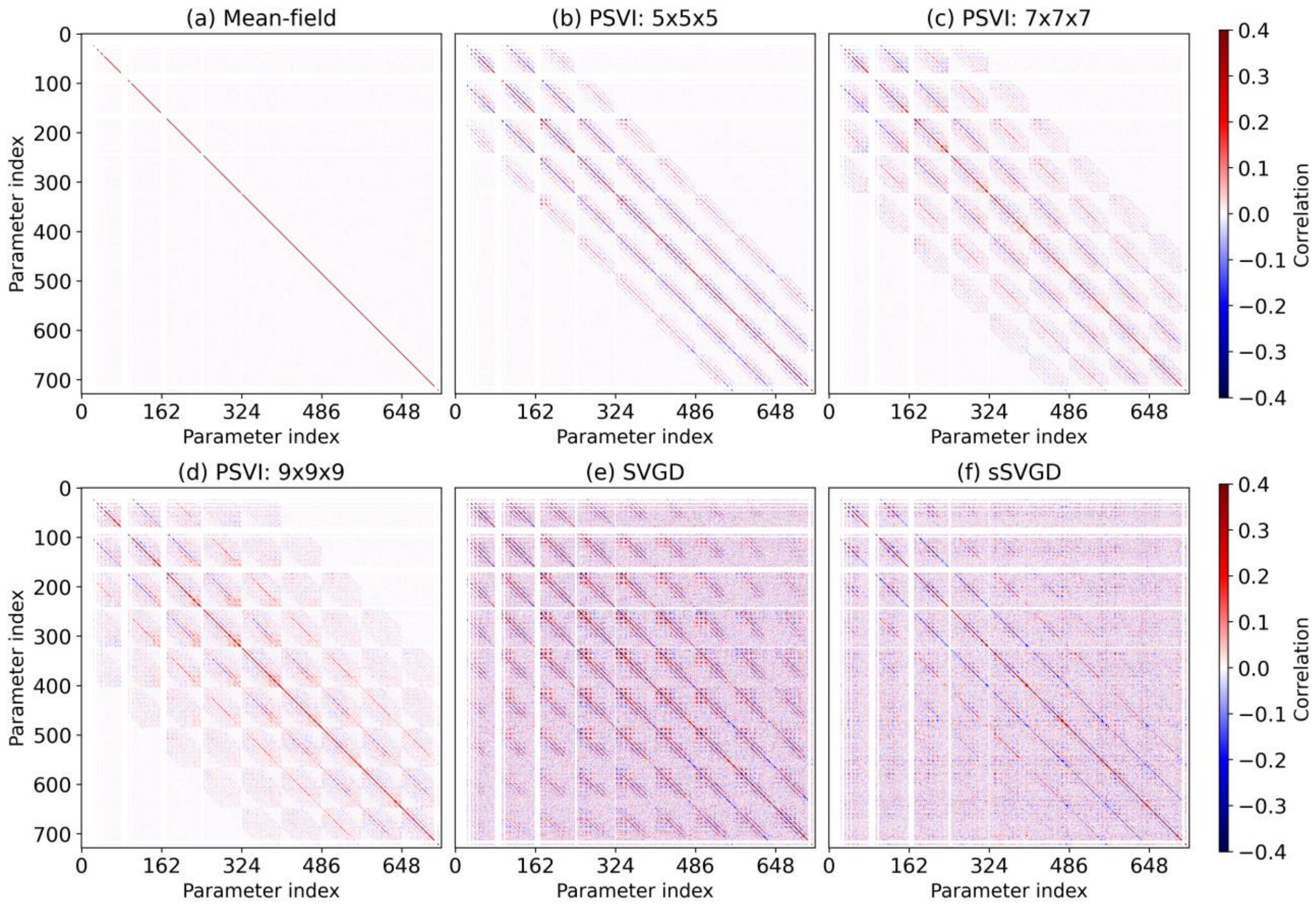

**Figure 16.** Correlation matrices for the subregion within the black box, spanning from the permeability cell located at (100m, 100m, 1020m) to (1000m, 1000m, 1200m) as shown in Figure 11b, computed from posterior samples using (a) mean-field ADVI, (b) PSVI with a 5×5×5 kernel, (c) PSVI with a 7×7×7 kernel, (d) PSVI with a 9×9×9 kernel, (e) SVGD, and (f) sSVGD.

The correlation structure from sSVGD, shown in Figure 16f, is similar to those obtained using PSVI. One subtle difference is the presence of a very weak blue off-diagonal line, located 243 indices away from the diagonal, indicating a weak correlation between each permeability cell and those located three layers above and below it. However, since the magnitude of this correlation is extremely small and none of the PSVI methods capture it, we conclude that a 5×5×5 correlation kernel in PSVI is sufficient to capture the essential correlation structure among permeability parameters.

Another notable result is the correlation matrix computed from the SVGD posterior samples, shown in Figure 16e. This suggests that each permeability cell is correlated with cells located six rows above or below, as well as with cells six columns away. This correlation pattern appears unreasonable as it is inconsistent with all other correlation matrices obtained from PSVI and sSVGD. This behaviour is likely due to the limited number of samples used in SVGD to approximate a high-dimensional posterior distribution, which can introduce spurious correlations. This effect is explored further in the Discussion below.

To further evaluate the performance of the posterior pdf approximation using different methods, we present two vertical permeability profiles in Figure 17, corresponding to the two red triangles shown in Figure 11b, starting from cells located at (950m, 950m, 1080m) and (550m, 550m, 1020m). The mean permeability estimates shown by the black lines generally resemble the true permeability values shown by the red lines. From the first row, we can see that all the true permeability values fall within the range of medium to high probability indicated by the more yellow regions. This demonstrates that all six methods provide a posterior approximation that successfully captures the true permeability within the posterior marginal distributions.

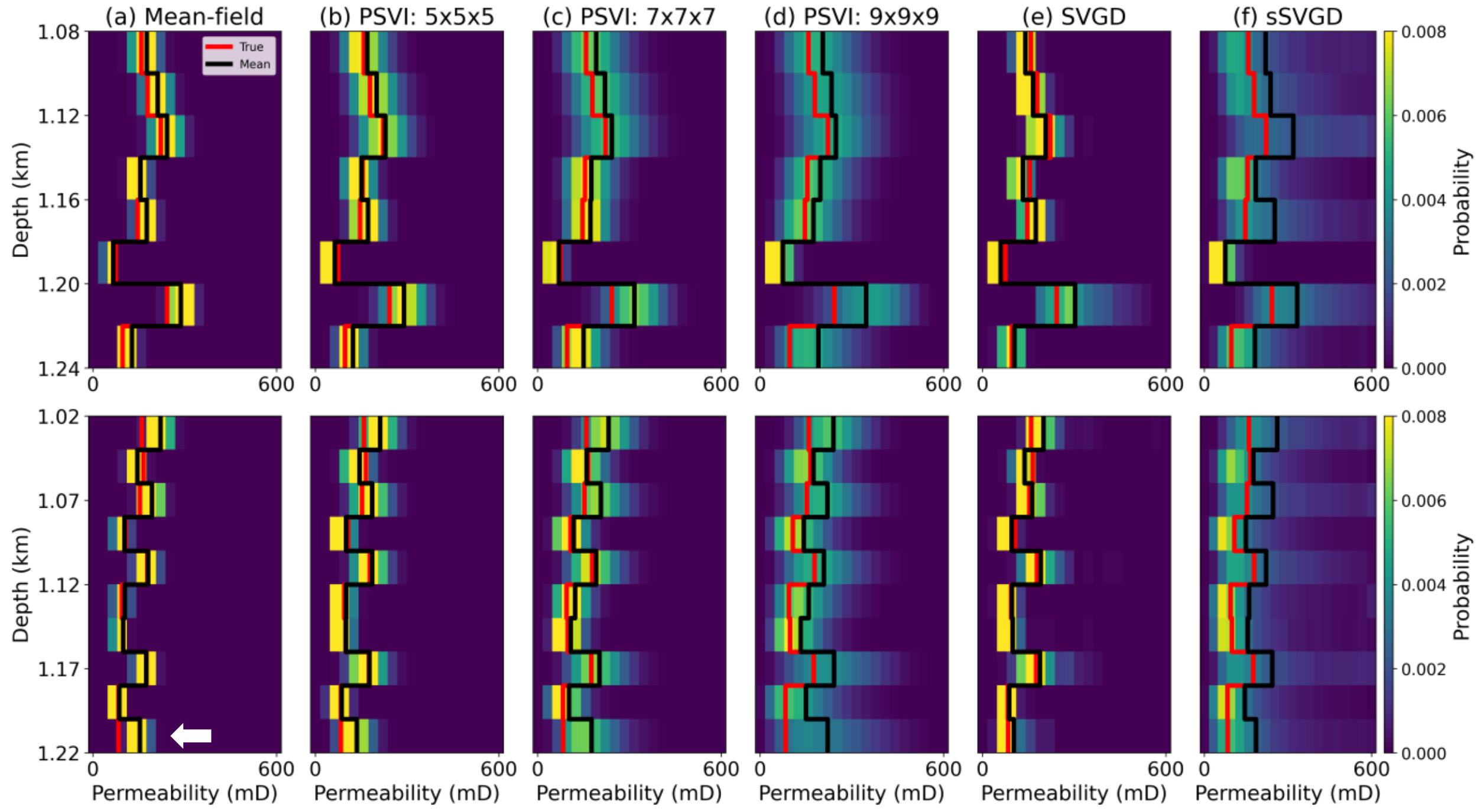

**Figure 17.** Marginal posterior distributions along two vertical profiles, indicated by the red triangles in Figure 11b, starting at permeability parameters (950m, 950m, 1080m) and (550m, 550m, 1020m). Yellow regions indicate areas of high posterior probability, while blue regions represent areas of zero posterior probability. The red line shows the true permeability values, and the black line shows the mean permeability values along the vertical profiles. The white arrows highlight regions where the true permeability lies in areas of low posterior probability.

However, as shown in the second row for mean-field ADVI, the true permeability value between 1.20 km and 1.22 km is assigned a very low posterior probability, as indicated by the white arrow. This is undesirable, as it indicates that the posterior pdf almost fails to encompass all plausible permeability values that can explain the observed data. By contrast, for the PSVI methods with the three different correlation kernel sizes, all true permeability values fall within intermediate to high probabilities, demonstrating improved performance over mean-field ADVI by better capturing key correlations.

Similarly to discussions above, the posterior range from PSVI remains narrower than that from sSVGD due to the Gaussian-shaped posterior assumption inherent in ADVI-based methods. However, SVGD is also free from the Gaussian-shaped posterior assumption, yet it also shows a much narrower posterior uncertainties, again likely resulting from mode collapse.

Similarly to the comparisons of computational efficiency above, mean-field ADVI and PSVI were run for 6,000 iterations with two Monte Carlo samples per iteration, giving 24,000 simulation runs for each method. Full-rank ADVI was run for 6,000 iterations with up to twenty Monte Carlo samples per iteration, giving 240,000 simulation runs; however, achieving stable results comparable in quality to the other methods would likely require additional iterations and/or more Monte Carlo samples per iteration. SVGD used 300 particles over 800 iterations, resulting in 480,000 simulations, while sSVGD used 20 chains with 8,000 iterations each, resulting in 320,000 simulations.

We also note that the dimensionality of the 2D case is 400, whereas the 3D case has 3,693 parameters, an increase of about ninefold. Despite this, and with the exception of full-rank ADVI which becomes difficult to converge in the 3D case, the number of simulations required by the variational inference methods increases by less than a factor of two. This suggests that these variational inference methods scale well with increasing dimensionality.

Table 2. The computational resources used in each inversion for the 3D test.

| Inversion methods | Number of simulations | Wall-clock time | Number of CPUs |
|---|---|---|---|
| Mean-field ADVI | 24,000 | 25.29 hours | 2 |
| PSVI | 24,000 | 27.88 hours | 2 |
| Full-rank ADVI | > 240,000 | 28.88 hours | 20 |
| sSVGD | 320,000 | 59.95 hours | 20 |
| SVGD | 480,000 | 105.91 hours | 20 |

We observe that mean-field ADVI and PSVI are the most computationally efficient methods, while PSVI outperforms mean-field ADVI in terms of the accuracy of its posterior pdf approximation by capturing the significant correlations among permeabilities. By contrast, optimising full-rank ADVI is significantly more challenging and computationally expensive in high-dimensional settings compared to mean-field ADVI, PSVI and potentially even sSVGD, as it still did not converge after 240,000 forward evaluations. This indicates that full-rank ADVI is not well suited for high-dimensional problems, since the size of the covariance matrix grows too rapidly with dimensionality $n$.

SVGD and sSVGD are not constrained by the Gaussian-shaped posterior assumption and in principle can provide more accurate posterior pdf approximations. However, the mode collapse problem inherent in SVGD makes it more challenging to apply in high-dimensional inverse problems. To avoid mode collapse, it may be necessary to significantly increase the number of particles to fully span the posterior space and maintain reasonable particle distances. As shown in Table 2, the computational cost of SVGD and sSVGD is already considerably higher than that of mean-field ADVI and PSVI, and increasing the number of particles for SVGD would make it even more computationally expensive.

## 4 Discussion

While PSVI provides the best trade-off between performance and efficiency in the above examples, the use of particle-based methods such as SVGD and sSVGD (or other Monte Carlo methods, e.g., Betancourt, 2018) may be justified in applications where accurate posterior estimation is critical and substantial computational resources are available. For example, when modelling leakage risk in nuclear waste disposal, or assessing the risk of fracturing the caprock due to pressure build-up in $CO_2$ sequestration, accurate uncertainties estimates are important, potentially making sSVGD and SVGD suitable choices, at least to obtain final results. For applications such as tracking the movement of the $CO_2$ plume during sequestration or when performing initial exploratory tests for more critical applications such as nuclear waste storage, computational efficiency may be more important and reasonable accuracy sufficient, so a less costly method such as PSVI would be more appropriate (Zhao & Curtis, 2025c).

The choice of kernel size is important in PSVI because it determines which posterior correlations can be represented in the variational covariance. To clarify the consequences of kernel mis-specification, it is useful to view mean-field ADVI (diagonal covariance) and full-rank ADVI (dense covariance) as two extreme cases of PSVI. If the kernel is chosen to be too small relative to the true posterior dependence structure, PSVI cannot represent important correlations and the approximation moves toward the mean-field assumption of independent parameters. In that case, posterior uncertainty is typically underestimated (Zhang et al., 2023) and this behaviour is visible in both the 2D and 3D examples: in Figures 10 and 17, the posterior ranges become overly narrow and can exclude the true model parameters. When this occurs, the variational posterior pdf does not cover the full set of plausible solutions that can explain the observed data, which is undesirable for uncertainty quantification. On the other hand, if the kernel is larger than necessary, PSVI can still capture the relevant correlations and the posterior estimate usually changes only marginally, but convergence to the solution can become numerically unstable. In the 2D case, a $5 \times 5$ kernel is sufficient, while full-rank ADVI allows correlations across the entire $20 \times 20$ permeability grid. In the 3D case we also tested larger kernels ($7 \times 7 \times 7$ and $9 \times 9 \times 9$) in the 3D case, and as shown in Figure 12 the results are very similar between the $5 \times 5 \times 5$ kernel and these larger choices. This indicates that once the kernel is large enough to include the dominant correlations, further enlargement yields diminishing returns. Larger kernels increase the number of variational parameters substantially and can make optimisation impractical in high dimensions as demonstrated in the 3D case (Figure 14), where increasing kernel size leads to a sharp rise in computational cost and optimisation instability.

To apply PSVI in other hydrogeological settings, we suggest selecting the kernel size by balancing representational adequacy and computational feasibility. If full-rank ADVI is computationally affordable and can be optimised robustly, it offers the most flexible covariance among ADVI variants. When full-rank inference is too costly or is unstable, PSVI provides a practical alternative. A sensible workflow is to start with a kernel that is expected to be more than sufficiently large, then adjust based on feasibility. Kernel design should also reflect expected physical behaviour and anisotropy. For example, our results indicate stronger vertical than horizontal correlations (Figures 6 and 16), consistent with $CO_2$ migration patterns, and this could be reflected in the kernel shape if necessary.

PSVI and full-rank ADVI approximate the posterior pdf using a Gaussian variational family in the adopted parametrisation, which yields a unimodal approximation. This assumption can be restrictive in strongly non-Gaussian geological settings. For example, in facies-based or object-based channel models the joint posterior pdf can become multimodal due to alternative connectivity configurations that explain the data equally well. A Gaussian variational family cannot represent such multimodality or sharp interfaces, and may therefore over-smooth the inferred fields and under-represent uncertainty. A natural extension is to adopt a more expressive non-Gaussian variational family to approximate the posterior pdf. For example, Gaussian mixture variational distributions (Locatello et al., 2018; Shahraeeni et al., 2012; Zhao & Curtis, 2024a), or other flexible families such as normalising flows (Papamakarios et al., 2021; Rakoczi et al., 2023; Rezende & Mohamed, 2015; Zhao et al., 2022), can capture multimodality and non-elliptical posterior structures more effectively than a single Gaussian, at the cost of additional computational complexity.

The mode collapse observed in our SVGD runs should not be interpreted as a universal failure of SVGD, but rather as the outcome of applying SVGD under a constrained configuration in a high-dimensional setting. Two factors contribute to this poor performance. First, SVGD can become challenging in high dimensions because commonly used kernels may lose effective particle interactions (for any computationally feasible number of particles, all particles are far apart in high-dimensional settings). This renders the interaction term in equation (D3) ineffective, allowing the convergence term to cause a collapse of all particles towards a single region of parameter space. Second, practical choices such as particle count and step size strongly influence performance even in lower-dimensional problems. With a limited number of particles, SVGD may fail to resolve multimodality or elongated posterior structure. Our experiments support this interpretation: in the 2D example, where the particle count is comparable to the problem dimensionality, SVGD produces stable and informative posterior pdf approximations. In the 3D example, where the particle count is much smaller than the dimensionality of the parameter space, SVGD exhibits mode collapse. SVGD therefore remains attractive when a flexible, non-Gaussian approximation is required and a sufficiently large particle ensemble is computationally feasible. Accordingly, our conclusions regarding SVGD are specific to the particle budgets and parametrisation choices tested in this high-dimensional setting, rather than a general statement that SVGD is unsuitable for geophysical or fluid flow inference problems.

The SVGD results in Figure 16e appear to show that each permeability cell is correlated with cells located six rows above or below, as well as with cells six columns away laterally – neither of which was verified by any other method. This can be explained by the fact that these correlations were estimated using fewer samples than the number of dimensionality (parameters), as illustrated in Appendix F.

In this work we have not included spatial correlations that may be present in seismic velocity data. Seismic velocity is typically estimated by tomography methods, using information in the kinematics (travel time tomography) or dynamics (waveform inversion) of seismic waves, and the resulting inverted velocity models tend to exhibit spatial correlation between nearby locations (Zhang & Curtis, 2021; Zhao & Curtis, 2024b). Incorporating the correlation derived from seismic tomography could potentially further enhance local trade-offs and reveal stronger interactions among permeability parameters that are spatially close, particularly since the correlations observed herein appear to have the same signs as those obtained in the tomographic studies cited above.

In real-world reservoir modelling, fluid flow simulations are usually computationally intensive, with a single forward run potentially taking several hours. This challenge becomes even more pronounced in inverse problems, where thousands or millions of simulations may be needed to estimate posterior distributions reliably. Even the most computationally efficient methods examined in this study, such as ADVI, may require days or weeks to converge when applied to field-scale models. In this context, incorporating machine learning based surrogate models might be justified. However, unlike standard machine learning applications, such as natural language processing where large datasets are readily available, training surrogate models for reservoir simulations demands carefully selected samples from the parameter space, each costly to generate (Guo et al., 2023; Wen et al., 2021). Despite these challenges, hybrid approaches that combine inversion methods with the training of machine learning surrogates hold important potential to make Bayesian inference feasible for large-scale reservoir problems and warrant further investigation (Fu et al., 2025; Sun et al., 2023).

In this work we use seismic velocity as the seismic-state variable, rather than working directly with seismic trace data. It would be straightforward to couple the current framework to a post-stack, trace-by-trace convolutional forward model for normal-incidence data, provided that the rock-physics model is formulated to predict impedance-related properties. Such a formulation is computationally inexpensive and may be appropriate for amplitude-based studies. However, the convolutional model does not capture the full physics of wave propagation exploited by more sophisticated waveform-based imaging algorithms. Our use of velocity as the seismic-state variable is motivated by the fact that the coupled workflow is intended to be compatible with most types of seismic inversion, such as travel-time tomography and full waveform inversion, for which the primary unknown is a velocity (or elastic) model, and the forward function consists of simulating wave propagation potentially followed by signal processing (e.g., noise removal or travel-time picking). The velocity-based formulation adopted here provides a direct route to coupling with any seismic imaging approaches within the same parametrisation.

A more complete rock-physics model would allow the dry frame bulk $B_{\mathrm{frame}}$ and shear modulus $\mu$ to vary with porosity, since both $B_{\mathrm{frame}}$ and $\mu$ depend on grain packing, cementation, and porosity, and are therefore expected to change spatially as porosity $\phi$ varies. In this work, we adopt constant values for $B_{\mathrm{frame}}$ and $\mu$ as a simplifying approximation to keep the patchy-saturation velocity mapping computationally inexpensive and to focus the study on the variational inference methodology and its coupling with multiphase flow. We acknowledge that, for the porosity range implied by equation (17), a fixed value such as $B_{\mathrm{frame}} \approx 2.56$ GPa may not be representative, and that a porosity-dependent dry-frame model $B_{\mathrm{frame}}(\phi)$ and $\mu(\phi)$ would be more consistent (Liu & Grana, 2018). Incorporating such a relation, for example using a calibrated empirical trend or standard stiff sand model (Dvorkin et al., 2014), is straightforward within the present workflow and is an important direction for future work aimed at improved quantitative realism.

Variational autoencoders and generative adversarial networks (GANs) can also be trained to incorporate geological knowledge (Bloem & Curtis, 2024; Jiang & Jafarpour, 2021; Laloy et al., 2018; Miele & Linde, 2025). The prior samples generated at each update step of inference methods can then be guided by the geological information embedded in these trained neural networks. Additionally, these neural networks can be used to reduce the dimensionality of the model parameters to be estimated, which can accelerate the convergence rate of inference methods. More recently, diffusion models have been proposed as a class of generative models (Heek et al., 2024; Ho et al., 2020; Peebles et al., 2022; Song et al., 2020) and applied to inference problems (Fan et al., 2024; Feng et al., 2025; Wang et al., 2023). For example, Fan et al. (2025) recast the typically unsupervised diffusion-generation procedure into a supervised learning framework by first generating a large synthetic labelled dataset using a conditional generative model: they approximate the gradient of the posterior pdf with a mini-batch Monte Carlo estimator and integrate a reverse-time ordinary differential equation (ODE) to obtain samples. They then train a supervised neural network to rapidly predict $CO_2$ saturation and pressure fields from monitoring data. In contrast to these amortised, simulation-trained approaches that learn a conditional sampler or inference network across many possible observations, PSVI is a non-amortised variational inversion method that directly optimises an approximate posterior pdf for each observed dataset, avoiding the need for large offline training sets. That said, prior information learned by a diffusion model can also be incorporated within variational frameworks such as PSVI, for example as an informative prior or an initializer, to accelerate convergence.

In this work, we assign a uniform distribution to the model parameters to represent weak prior knowledge. However, in real applications, prior information is often more complex. In such cases, it may be possible to design a more expressive covariance structure within PSVI to better capture realistic correlations among model parameters. Prior replacement methods can also be used to efficiently explore the effects of different prior information on the inversion results (Zhao & Curtis, 2024c).

In principle it is possible to design physically structured SVGD and sSVGD methods that both capture reliable uncertainties estimates and improve computational efficiency. For example, Chen & Ghattas (2020) introduced projected SVGD, which reduces the high-dimensional parameter space to a lower-dimensional subspace capturing the components that most significantly shift the prior toward the posterior distributions. This approach helps mitigate the mode collapse problem, as the repulsive force among particles remains strong within the lower-dimensional space. Building on this idea, further extensions could involve projecting sSVGD into a low-dimensional subspace as well, potentially accelerating convergence rates.

Overall, based on the results herein we conclude that either PSVI or sSVGD should be used for Bayesian estimation of reservoir properties. The latter method should only be chosen if the dimensionality (number of parameters) is relatively low compared to typical reservoir simulation models, and if the need for accurate risk estimates over-rides considerations of computational cost.

## 5 Conclusions

This study evaluated the performance of different variational inference methods, specifically mean-field ADVI, PSVI, full-rank ADVI, SVGD and sSVGD, for solving dynamic flow inverse problems. The results demonstrate clear trade-off between computational efficiency and inversion accuracy across these methods. PSVI provides a promising balance between accuracy and efficiency by capturing important parameter correlations while maintaining a computational cost comparable to the simplest method – mean-field ADVI. Although full-rank ADVI can in theory describe full model correlations among parameters, it is extremely expensive (if not impossible) to implement for high dimensional problems in terms of both computational cost and memory requirements. Particle-based methods such as SVGD and sSVGD can represent non-Gaussian posterior structures more flexibly than a single Gaussian variational family, because they do not rely on a closed-form, analytic variational family of distributions. However, these methods come with significantly higher computational cost. SVGD also exhibits mode collapse and introduced spurious correlations in high-dimensional problems when the number of particles was limited, particularly when it was smaller than the dimensionality of the space.

## Acknowledgments

The authors wish to thank the sponsors of the Edinburgh Imaging Project (EIP), TotalEnergies and BP, for supporting this research. We also gratefully acknowledge the valuable discussions with the developers of JutulDarcy (Moyner, 2024), JutulDarcyRules (Louboutin et al., 2023), JUDI (Witte et al., 2019), and VIP (Zhang & Curtis, 2023b).

## Open Research

Software used for the variational methods as well as the McMC can be found at PyMC3 website (https://docs.pymc.io/en/v3/, Salvatier et al., 2016). Software used to perform Automatic Differentiation can be found at PyTorch website (https://pytorch.org/, Paszke et al., 2019).

## Conflict of Interest Disclosure

The authors declare there are no conflicts of interest for this manuscript.

**APPENDIX A: Gradients for ADVI**

The Evidence Lower Bound (ELBO) is expressed as

$$\mathrm{ELBO} = E_{N(\boldsymbol{\eta}|\mathbf{0},\mathbf{I})}\left[\log p\left(T^{-1}\left(R_{\boldsymbol{\varphi}}^{-1}(\boldsymbol{\eta})\right), \mathbf{d}_{obs}\right)\right] + \log\left|\det \mathbf{J}_{T^{-1}}\left(R_{\boldsymbol{\varphi}}^{-1}(\boldsymbol{\eta})\right)\right| + H[q(\boldsymbol{\mathcal{S}};\boldsymbol{\varphi})] \qquad (A1)$$

where the entropy term $H[q(\boldsymbol{\mathcal{S}};\boldsymbol{\varphi})] = -E_q[\log q(\boldsymbol{\mathcal{S}};\boldsymbol{\varphi})]$, and the variational parameters are denoted by $\boldsymbol{\varphi} = \{\boldsymbol{\mu}, \mathbf{L}\}$. By applying the chain rule, the gradient of the ELBO with respect to $\boldsymbol{\mu}$ can be computed as

$$\nabla_{\boldsymbol{\mu}}\mathrm{ELBO} = \nabla_{\boldsymbol{\mu}}\left\{E_{N(\boldsymbol{\eta}|\mathbf{0},\mathbf{I})}\begin{bmatrix}\log p\left(T^{-1}\left(R_{\boldsymbol{\varphi}}^{-1}(\boldsymbol{\eta})\right), \mathbf{d}_{obs}\right) \\ +\log\left|\det \mathbf{J}_{T^{-1}}\left(R_{\boldsymbol{\varphi}}^{-1}(\boldsymbol{\eta})\right)\right|\end{bmatrix} + H[q(\boldsymbol{\mathcal{S}};\boldsymbol{\varphi})]\right\} \qquad (A2)$$

Using the Dominated Convergence Theorem (Çinlar, 2011), we can exchange the gradient and the expectation

$$\nabla_{\boldsymbol{\mu}}\mathrm{ELBO} = E_{N(\boldsymbol{\eta}|\mathbf{0},\mathbf{I})}\{\nabla_{\boldsymbol{\mu}}[\log p\left(T^{-1}\left(R_{\boldsymbol{\varphi}}^{-1}(\boldsymbol{\eta})\right), \mathbf{d}_{obs}\right)] + \nabla_{\boldsymbol{\mu}}(\log\left|\det \mathbf{J}_{T^{-1}}\left(R_{\boldsymbol{\varphi}}^{-1}(\boldsymbol{\eta})\right)\right|)\} \qquad (A3)$$

Note that the entropy term $H$ does not depend on $\boldsymbol{\mu}$, and therefore its gradient vanishes.

Applying the chain rule

$$\nabla_{\boldsymbol{\mu}}\mathrm{ELBO} = E_{N(\boldsymbol{\eta}|\mathbf{0},\mathbf{I})}\begin{bmatrix}\nabla_{\mathbf{m}}\log p(\mathbf{m}, \mathbf{d}_{obs})\nabla_{\boldsymbol{\mathcal{S}}}T^{-1}(\boldsymbol{\mathcal{S}})\nabla_{\boldsymbol{\mu}}R_{\boldsymbol{\varphi}}^{-1}(\boldsymbol{\eta}) \\ +\nabla_{\boldsymbol{\mathcal{S}}}\log|\det \mathbf{J}_{T^{-1}}(\boldsymbol{\mathcal{S}})|\,\nabla_{\boldsymbol{\mu}}R_{\boldsymbol{\varphi}}^{-1}(\boldsymbol{\eta})\end{bmatrix} \qquad (A4)$$

which simplifies to

$$\nabla_{\boldsymbol{\mu}}\mathrm{ELBO} = E_{N(\boldsymbol{\eta}|\mathbf{0},\mathbf{I})}[\nabla_{\mathbf{m}}\log p(\mathbf{m}, \mathbf{d}_{obs})\nabla_{\boldsymbol{\mathcal{S}}}T^{-1}(\boldsymbol{\mathcal{S}}) + \nabla_{\boldsymbol{\mathcal{S}}}\log|\det \mathbf{J}_{T^{-1}}(\boldsymbol{\mathcal{S}})|] \qquad (A5)$$

Similarly, the gradient of the ELBO with respect to $\mathbf{L}$ is

$$\nabla_{\mathbf{L}}\text{ELBO} = \nabla_{\mathbf{L}}\{E_{N(\boldsymbol{\eta}|\mathbf{0},\mathbf{I})}\begin{bmatrix}\log p\left(T^{-1}\left(R_{\boldsymbol{\varphi}}^{-1}(\boldsymbol{\eta})\right),\mathbf{d}_{obs}\right)\\ +\log\left|\det\mathbf{J}_{T^{-1}}\left(R_{\boldsymbol{\varphi}}^{-1}(\boldsymbol{\eta})\right)\right|\end{bmatrix} + \frac{k}{2} + \frac{k}{2}\log(2\pi) + \frac{1}{2}\log|\det(\mathbf{L}\mathbf{L}^T)|\} \quad (A6)$$

Using the Dominated Convergence Theorem to exchange the gradient and expectation

$$\nabla_{\mathbf{L}}\text{ELBO} = E_{N(\boldsymbol{\eta}|\mathbf{0},\mathbf{I})}\begin{bmatrix}\nabla_{\mathbf{L}}\left\{\log p\left(T^{-1}\left(R_{\boldsymbol{\varphi}}^{-1}(\boldsymbol{\eta})\right),\mathbf{d}_{obs}\right)\right\}\\ +\nabla_{\mathbf{L}}\left(\log\left|\det\mathbf{J}_{T^{-1}}\left(R_{\boldsymbol{\varphi}}^{-1}(\boldsymbol{\eta})\right)\right|\right)\end{bmatrix} + \nabla_{\mathbf{L}}\frac{1}{2}\log|\det(\mathbf{L}\mathbf{L}^T)| \quad (A7)$$

Applying the chain rule

$$\nabla_{\mathbf{L}}\text{ELBO} = E_{N(\boldsymbol{\eta}|\mathbf{0},\mathbf{I})}\begin{bmatrix}\nabla_{\mathbf{m}}\log p(\mathbf{m},\mathbf{d}_{obs})\nabla_{\boldsymbol{\mathcal{S}}}T^{-1}(\boldsymbol{\mathcal{S}})\nabla_{\mathbf{L}}R_{\boldsymbol{\varphi}}^{-1}(\boldsymbol{\eta})\\ +\nabla_{\boldsymbol{\mathcal{S}}}\log|\det\mathbf{J}_{T^{-1}}(\boldsymbol{\mathcal{S}})|\,\nabla_{\mathbf{L}}R_{\boldsymbol{\varphi}}^{-1}(\boldsymbol{\eta})\end{bmatrix} + (\mathbf{L}^{-1})^T \quad (A8)$$

which simplifies to

$$\nabla_{\mathbf{L}}\text{ELBO} = E_{N(\boldsymbol{\eta}|\mathbf{0},\mathbf{I})}\begin{bmatrix}(\nabla_{\mathbf{m}}\log p(\mathbf{m},\mathbf{d}_{obs})\nabla_{\boldsymbol{\mathcal{S}}}T^{-1}(\boldsymbol{\mathcal{S}})\\ +\nabla_{\boldsymbol{\mathcal{S}}}\log|\det\mathbf{J}_{T^{-1}}(\boldsymbol{\mathcal{S}})|)\boldsymbol{\eta}^T\end{bmatrix} + (\mathbf{L}^{-1})^T \quad (A9)$$

**APPENDIX B: Governing Equations**

Here we describe the governing equations for fluid flow and the rock physics model used to transform saturation into velocity data. For the fluid flow simulation, the initial pressure is set to hydrostatic conditions. The initial surface temperature is assumed to be constant at 23.5°C, increasing with a thermal gradient of 25°C per kilometre. Mass conservation for each component in fluid flow is expressed as

$$\frac{\partial}{\partial t}\left(\phi\left[S_w c_{iw}\rho_w + S_g c_{ig}\rho_g\right]\right) + \nabla\cdot\left(u_w c_{iw}\rho_w + u_g c_{ig}\rho_g + D_{iw}\rho_w + D_{ig}\rho_g\right) = Q_i, i = 1,\dots,n \quad (B1)$$

where $t$ is time, $\phi$ is porosity, $S_\eta$ and $\rho_\eta$ are the fluid saturation and density of phase $\eta$ (which can be $CO_2$ phase $g$ and water phase $w$). $u_\eta$ is the phase superficial velocity, and $Q_i$ is the phase sink/source term for component $i$, the total number of components is $n$. $c_{iw}$ and $c_{ig}$ are the mole fraction for component $i$ in water and gas phase, respectively. The saturation constraint is written

$$\sum_{i(\eta)}^{N_{phase}} S_i = 1 \qquad (B2)$$

where $N_{phase}$ is the number of phases, which in this study includes the water and $CO_2$ phases. The superficial phase velocity is modelled using the extended Darcy's law,

$$u_\eta = -\frac{k k_{r\eta}}{\mu_\eta} \nabla \Phi_\eta \qquad (B3)$$

where $k$ is the absolute permeability, $k_{r\eta}$ is the phase relative permeability for phase $\eta$. $\mu_\eta$ is the phase viscosity, and the phase potential $\Phi_\eta$ is given by

$$\Phi_\eta = P_\eta + \rho_\eta g z \qquad (B4)$$

where $P_\eta$ is the phase pressure, $\rho_\eta$ is the phase density, $g$ is the gravitational acceleration, and $z$ is the phase hydrostatic height.

The capillary pressure $P_c$ is modelled as the pressure difference between the $CO_2$ phase pressure $P_g$ and the water phase pressure $P_w$,

$$P_c = P_g - P_w \qquad (B5)$$

The values of parameters used in the fluid flow simulations are presented in Table B1.

Table B1. Parameters used in the flow simulations.

| Parameters | Values |
|---|---|
| Water relative permeability endpoint | 1.0 |
| $CO_2$ relative permeability endpoint | 0.95 |
| Water viscosity | 1 centipoise (cP) |
| $CO_2$ viscosity | 0.1 cP |
| Water density | 1190 kg/m$^3$ |
| $CO_2$ density | 700 kg/m$^3$ |
| Residual water saturation | 0.2 |
| Residual $CO_2$ saturation | 0 |
| $CO_2$ injection rate | 1 Mt/year |

After obtaining the $CO_2$ saturation from the fluid flow simulation, the patchy saturation model is employed to predict seismic velocity (Dhananjay, 2006; Gassmann, 1951). This model is applicable when the fluid distribution forms distinct patches that are significantly larger than individual pore spaces. Under this assumption, the bulk modulus $B_{rg}$ of the rock fully saturated with $CO_2$ is determined by solving the following equation:

$$\frac{B_{rg}}{B_{grain} - B_{rg}} = \frac{B_{rw}}{B_{grain} - B_{rw}} - \frac{B_w}{\phi(B_{grain} - B_w)} + \frac{B_g}{\phi(B_{grain} - B_g)} \qquad (B6)$$

where $B_{grain}$ denotes the bulk modulus of the rock grains, while $B_w$ and $B_g$ represent the bulk moduli of water and $CO_2$, respectively. $\phi$ refers to the porosity of the rock. $B_{rw}$ and $B_{rg}$ are the bulk moduli of the rock fully saturated with water and $CO_2$, respectively. The $B_{rw}$ can be calculated using

$$B_{rw} = B_{frame} + \frac{(1 - \frac{B_{frame}}{B_{grain}})^2}{\frac{\phi}{B_w} + \frac{1-\phi}{B_{grain}} - \frac{B_{frame}}{{B_{grain}}^2}} \qquad (B7)$$

where $B_{frame}$ denotes the bulk modulus of the rock frame. The bulk modulus of the rock partially saturated with $CO_2$ and water, $B_r(S_g)$, can be calculated using

$$B_r(S_g) = \left[(1 - S_g)\left(B_{rw} + \frac{4}{3}\mu\right)^{-1} + S_g\left(B_{rg} + \frac{4}{3}\mu\right)^{-1}\right]^{-1} - \frac{4}{3}\mu \qquad (B8)$$

where $S_g$ represents the saturation of $CO_2$. Since the shear modulus of a fluid is zero, it is assumed that the rock's shear modulus $\mu$ remains unchanged when $CO_2$ replaces water. Here $\mu$ denotes the dry rock frame shear modulus. It can be obtained from an independent rock-physics calibration or derived from mineral properties and porosity using an appropriate dry-frame model. In this work, we treat $\mu$ as a known parameter.

The density of the rock partially saturated with $CO_2$ and water, $\rho_r(S_g)$, is expressed as

$$\rho_r(S_g) = \rho_{\text{grain}}(1-\phi) + \phi\left[(1-S_g)\rho_w + S_g\rho_g\right] \quad (B9)$$

where $\rho_r$, $\rho_{\text{grain}}$, $\rho_w$, and $\rho_g$ are the density of the rock, rock grain, water, and $CO_2$, respectively. The rock velocity partially saturated with $CO_2$ and water, $v_p(S_g)$, can be computed using the relation

$$B_r(S_\text{g}) = \rho_r(S_g)\left(v_p(S_g)^2 - \frac{4}{3}v_s(S_g)^2\right) \quad (B10)$$

Additionally, the shear modulus $\mu$ is given by

$$\mu = \rho_r(S_g)v_s(S_g)^2 \quad (B11)$$

Finally, the rock velocity partially saturated with $CO_2$ and water, $v_p(S_g)$, can be obtained:

$$v_p(S_g) = \sqrt{\frac{B_r(S_\text{g})}{\rho_r(S_g)} + \frac{4}{3}\frac{\mu}{\rho_r(S_g)}} \quad (B12)$$

The parameter values used in the rock physics modelling are presented in Table B2 and are adopted from the Sleipner field (Strutz & Curtis, 2024a). We note that a more realistic grain density is 2,650 kg/m$^3$, as suggested by Strutz & Curtis (2024b). Here we use 2100 kg/m$^3$ to illustrate the proposed workflow. Since the rock-physics model is employed deterministically, this choice primarily rescales $v_p$ through the ${\rho_r}^{-1/2}$ dependence and therefore should not change the conclusions of the paper.

Table B2. Parameters used in the rock physics modelling.

| Parameters | Values |
|---|---|
| Bulk modulus of the rock grains, $B_{grain}$ | 39.3 GPa |
| Bulk modulus of the rock frame, $B_{frame}$ | 2.56 GPa |
| Shear modulus of the rock frame, $\mu$ | 8.5 GPa |
| Bulk modulus of the water, $B_w$ | 2.31 GPa |
| Bulk modulus of the $CO_2$, $B_g$ | 0.08 GPa |
| Rock density, $\rho_{\mathrm{grain}}$ | 2100 kg/m$^3$ |

**APPENDIX C: Correlation Estimation Using Different Observed Data Types**

Since we use both seismic velocity and well pressure data as observations to invert the permeability models, it is important to understand the inversion behaviour when using only well pressure or only velocity data. In particular, we aim to examine how different types of observed data influence the correlation structure among permeability parameters.

First, we consider the case where only the well pressure data are used as observations to invert the permeability models. We apply full-rank ADVI, SVGD and sSVGD, as these methods provide a full-rank covariance matrix, allowing us to analyse the complete correlation structure among permeability parameters. The setup for each method follows that described for the 2D example in Section 3. We present the correlation matrices for the full permeability models.

We see that the correlation matrices in Figure C1 are noticeably different from the one shown in Figure 6. In Figure 6, the correlation matrix displays an alternating pattern of negative and positive correlations along the vertical direction as the distance between two permeability parameters increases. By contrast, Figure C1 shows that the correlations between permeability parameters are consistently negative along the vertical direction.

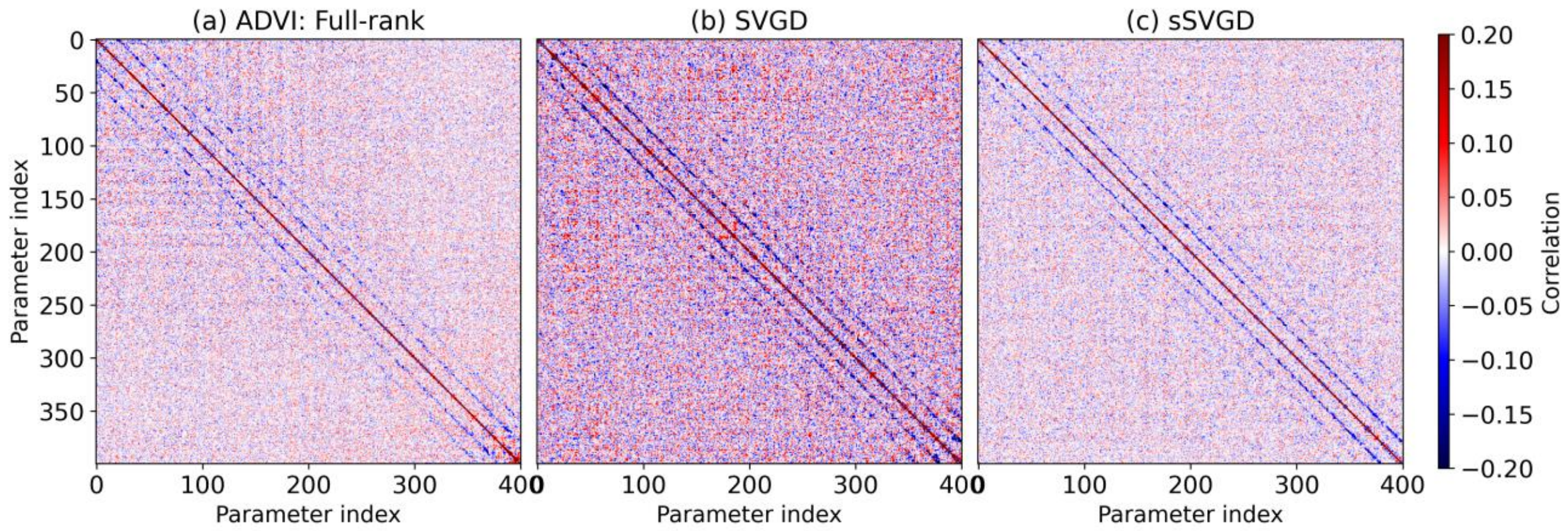

**Figure C1.** The correlation matrix computed using (a) full-rank ADVI, (b) SVGD, and (c) sSVGD, based on pressure data.

Since the observed well pressures are limited to sparse well locations, pressure must propagate between injector and monitoring wells across a significant portion of the domain. As a result, each pressure datum is influenced by permeability variations over a wide volume. This leads to global trade-offs during inversion: an increase in the permeability in one cell can be balanced by decreases in the permeability of a variety of other cells along the flow path such that the datum does not change, resulting in negative correlations shown in Figure C1.

Then we use only the velocity data as the observation to invert the permeability models. Again, we compute the correlation matrix to study the correlation among permeability parameters, as shown in Figure C2. We observe that the correlation structure is similar to that obtained when using both pressure and velocity data, as shown in Figure 6. Specifically, in the vertical direction, as the distance between two permeability parameters increases, the correlation is first negative and then positive.

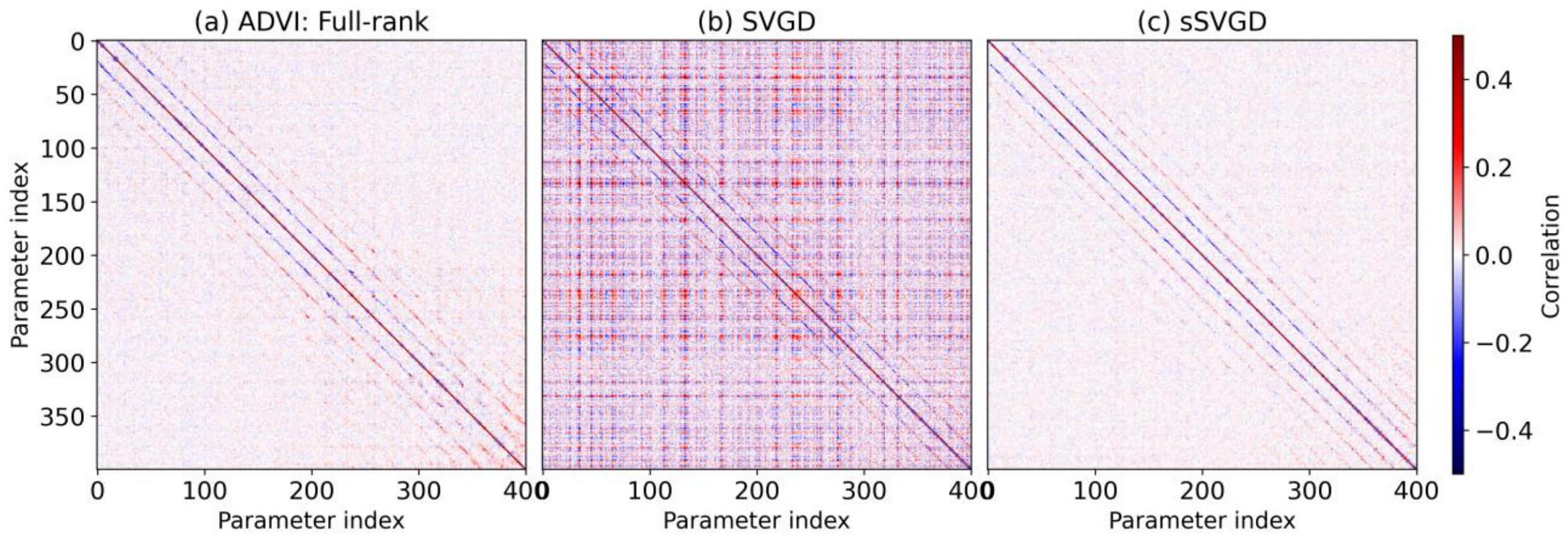

**Figure C2.** The correlation matrices computed using (a) full-rank ADVI, (b) SVGD, and (c) sSVGD, based on velocity data.

This phenomenon can be explained by the spatially comprehensive nature of the observed velocity data. A permeability increase in any particular cell may locally trade-off with permeability decreases in immediately neighbouring cells producing no net change in the observed datum in the particular cell, resulting in negative correlations in the inversion (similar to the mechanism described above for pressure data). However, there exists a velocity datum in every cell, and any negative change in a neighbouring cell will affect the velocity data in that cell unless it is balanced by a change in the opposite direction (now, a *positive* perturbation) in *its* neighbouring cells. This induces strong localisation in the sensitivity, and reverses its sign in each consecutive cell offset, creating a cascading adjustment and an alternating pattern of positive and negative correlations between each permeability cell and its neighbouring cells along the flow path.

Figures C1 and C2 indicate that both pressure and velocity data influence the permeability correlation structure. However, the overall correlation pattern among permeability parameters largely follows the structure derived from the velocity data, as seen in Figure 6. This is because pressure data are only available at a few specific well locations, whereas velocity data are available across the entire domain at multiple time steps. As a result, velocity data dominate the inversion over pressure data. We can also expect some degree of cancellation between the positive correlations generated by velocity data and the negative correlations generated by pressure data, but the positive correlations induced by velocity data are more pronounced and tend to dominate the combined correlation structure.

Nevertheless, whether the correlation between a permeability cell and its neighbouring cells is negative or positive, Figures C1 and C2 show that permeability parameters are only correlated with other cells at most two rows or two columns away. Therefore, using PSVI with a 5-by-5 correlation kernel should be sufficient to capture the posterior correlations.

**APPENDIX D: Mode Collapse of SVGD**

The mode collapse in SVGD can be understood through the update equation for each particle:

$$\boldsymbol{\phi}^{*}(\mathbf{m}) = \frac{1}{n}\sum_{j=1}^{n}\left[k(\mathbf{m}_i^l, \mathbf{m})\nabla_{\mathbf{m}_i^l}\log p(\mathbf{m}_i^l) + \nabla_{\mathbf{m}_i^l}k(\mathbf{m}_i^l, \mathbf{m})\right] \qquad (D1)$$

where $\mathbf{m}$ represents the model parameter, $\boldsymbol{\phi}^{*}(\mathbf{m})$ is the update direction for the particles, $n$ is the number of particles, and $\log p(\mathbf{m}_i^l)$ is the log target distribution of the particle $\mathbf{m}_i$ at iteration $l$. $k(\mathbf{m}_i^l, \mathbf{m})$ is a radial basis function (Liu & Wang, 2016), given by

$$k(\mathbf{m}_i^l, \mathbf{m}) = \exp\left(-\frac{1}{h}\left|\left|\mathbf{m}_i^l - \mathbf{m}\right|\right|^2\right) \qquad (D2)$$

where $h = \frac{d^2}{\log n}$, and $d$ is the median of pairwise particle distances. The gradient of the radial basis function is:

$$\nabla_{\mathbf{m}_i^l}k(\mathbf{m}_i^l, \mathbf{m}) = -\frac{2}{h}k(\mathbf{m}_i^l, \mathbf{m})(\mathbf{m}_i^l - \mathbf{m}) = -\frac{2}{h}\exp\left(-\frac{1}{h}\left|\left|\mathbf{m}_i^l - \mathbf{m}\right|\right|^2\right)(\mathbf{m}_i^l - \mathbf{m}) \qquad (D3)$$

The detailed derivation of equations D1 to D3 can be found in Liu & Wang (2016).

Equation (D1) shows the update equation for each particle in SVGD. The first term is known as the attraction or convergence term. When updating a particle $\mathbf{m}$, we compute the gradient $\nabla_{\mathbf{m}_i^l}\log p(\mathbf{m}_i^l)$ for all particles and combine their effects using the weighting coefficient $k(\mathbf{m}_i^l, \mathbf{m})$. When the distance between two particles is small and approaches zero, the kernel $k(\mathbf{m}_i^l, \mathbf{m})$ is close to one. This gives a large weight to the gradient of the nearby particle, meaning that the gradients from particles close to $\mathbf{m}$ have a strong influence on the update direction of $\mathbf{m}$. If the distance between two particles is large, the kernel $k(\mathbf{m}_i^l, \mathbf{m})$ is close to zero. This means the gradients from particles that are far away from $\mathbf{m}$ have little influence on its update direction.

The second term in equation D1 is known as the repulsive force, which is shown in equation D3. Let us first consider the update direction. If we compute the vector difference $\mathbf{m}_i^l - \mathbf{m}$, this vector points from $\mathbf{m}$ to $\mathbf{m}_i^l$. Since we have a negative sign in the gradient expression, the direction is reversed and points from $\mathbf{m}_i^l$ to $\mathbf{m}$. It is important to remember that the gradient we compute from equation D3 is applied in equation D1 to update the particle $\mathbf{m}$ instead of $\mathbf{m}_i^l$, so point from $\mathbf{m}_i^l$ to $\mathbf{m}$ will make $\mathbf{m}$ move away from $\mathbf{m}_i^l$, then it can be used to serve as a repulsive force.

We then understand the magnitude of the repulsive force by looking at how it depends on the distance between particles. When two particles are far apart, $k(\mathbf{m}_i^l, \mathbf{m})$ approaches zero, causing the repulsive force $\nabla_{\mathbf{m}_i^l} k(\mathbf{m}_i^l, \mathbf{m})$ to become negligible. This means that only a small repulsive force is applied when particles are already well separated. Conversely, when particles are close to each other, $k(\mathbf{m}_i^l, \mathbf{m})$ approaches one, and the resulting repulsive force $\nabla_{\mathbf{m}_i^l} k(\mathbf{m}_i^l, \mathbf{m})$ becomes large. This strong repulsion helps prevent particles from collapsing into the same location.

When we use a limited number of particles in a high-dimensional space, the relative distances between particles are large because particles end up in different corners of a hyper-cubic parameter space (Curtis & Lomax, 2001), causing the kernel values and repulsive force between particles to become small. As a result, the particles are mainly driven by the attraction term, which pulls them toward regions of high posterior density. This imbalance leads to mode collapse, where particles concentrate in a small region rather than spreading out to represent the full posterior distribution.

## APPENDIX E: Inversion Performance under Equal Forward Simulation Budgets

ADVI is run for 4,000 iterations with 2 Monte Carlo samples per iteration. SVGD uses 300 particles over 27 iterations, and sSVGD uses 20 chains with 400 iterations per chain. MH-McMC is run with 20 chains and 400 iterations. The corresponding inversion results are presented in Figures E1 and E2.

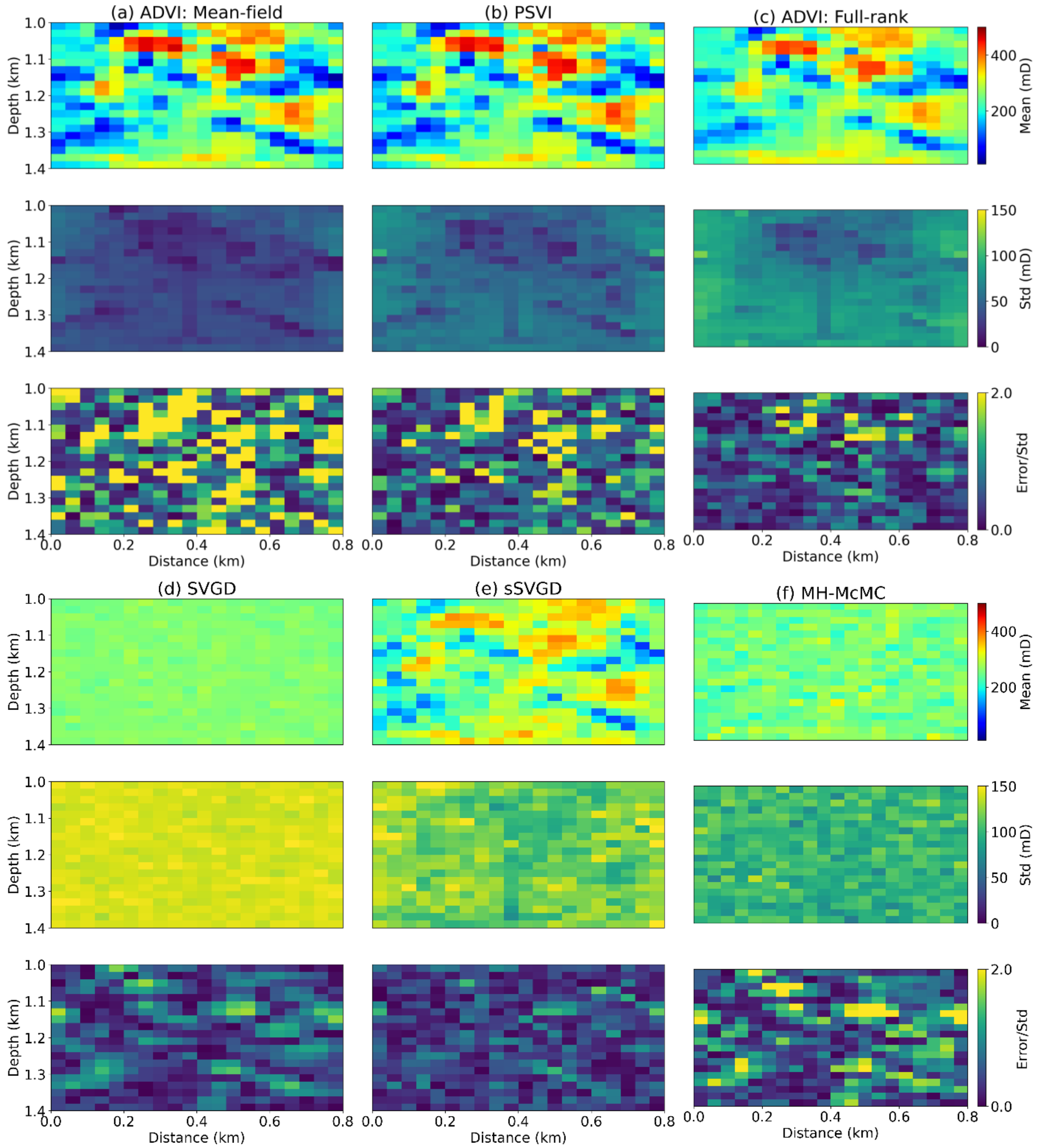

**Figure E1.** The mean (first row), standard deviation (second row), and relative error (third row), computed using (a) mean-field ADVI, (b) PSVI, (c) full-rank ADVI, (d) SVGD, (e) sSVGD, and (f) MH-McMC. All methods use 8,000 forward simulations.

The first row in Figure E1 shows that the mean permeability estimates from the ADVI methods are much closer to the true permeability models than those in the second row. The mean estimates from SVGD and MH-McMC lack any clear structural features and instead resemble the average of the uniform prior distribution, corresponding to the average value of the upper and lower bounds. sSVGD performs better than SVGD and MH-McMC but still falls far short of results from the ADVI methods when using the same number of forward simulations. The second row presents the standard deviation estimates, showing that SVGD, sSVGD and MH-McMC produce significantly higher uncertainties without clear structural patterns compared with the ADVI methods. The standard deviations from SVGD and sSVGD are on average even larger than those from MH-McMC, demonstrating a less informative posterior approximation.

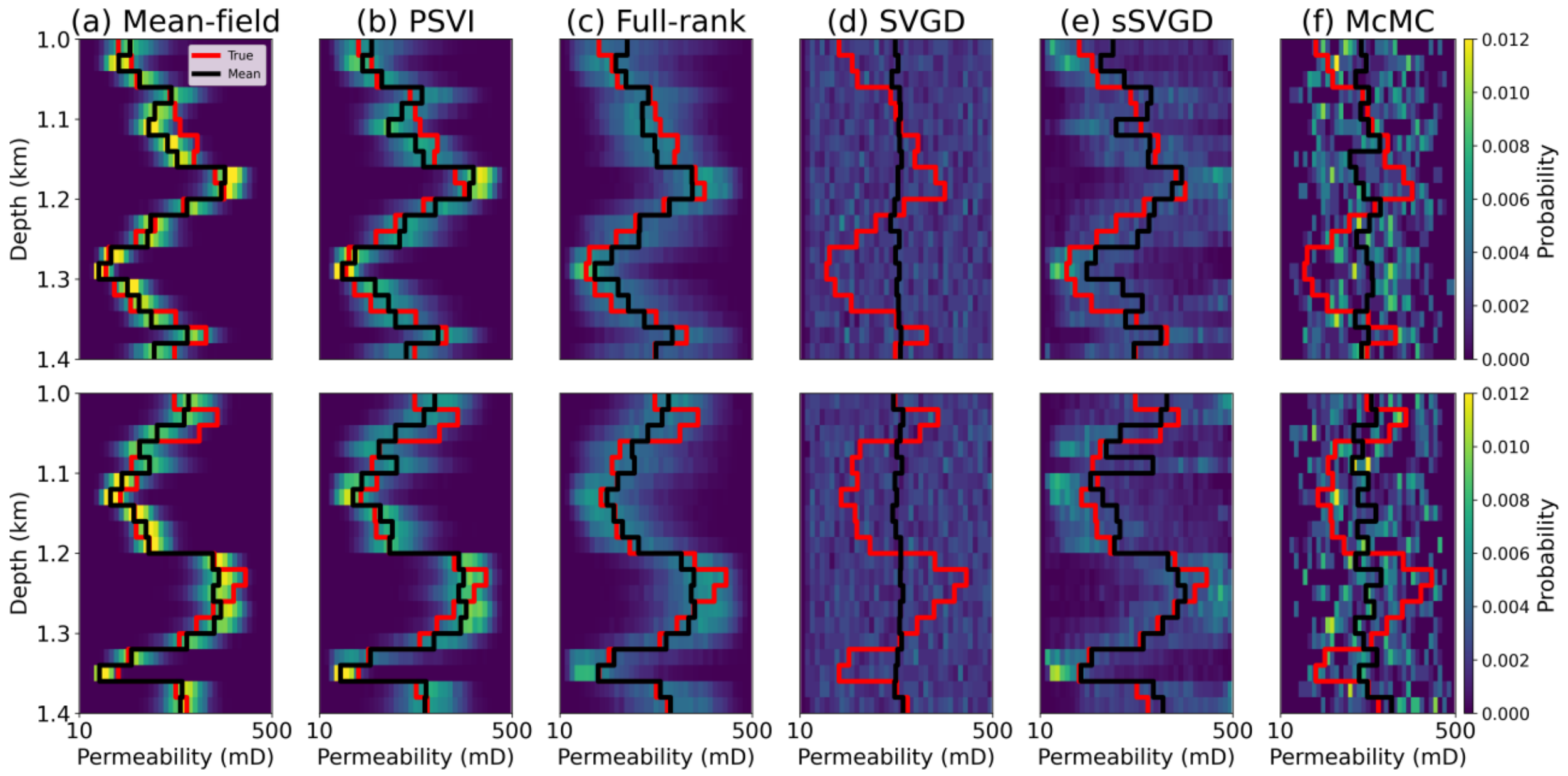

**Figure E2.** Marginal posterior distributions along two vertical profiles, starting at permeability parameters located at (140m, 1000m) and (700m, 1000m), computed using 8,000 forward simulations for each method. Yellow regions indicate areas of high posterior probability values, while blue regions represent

areas of zero probability value. The red line shows the true permeability values, and the black line indicates the mean permeability along the profiles.

Figure E2 shows the resulting marginal posterior distributions along two vertical profiles starting from locations (140m, 1000m) and (700m, 1000m). The posterior uncertainties computed from SVGD and MH-McMC spans nearly the entire range of permeability values, indicating limited updating of the prior distribution. The posterior range obtained from sSVGD shows noticeable improvement compared to SVGD and MH-McMC under the same number of forward simulations. This improvement is likely to be due to sSVGD performing more update iterations than SVGD, sharing the same iteration count as MH-McMC, yet benefiting from guidance by the Stein gradient (see Appendix D). These results highlight the inefficiency of SVGD and MH-McMC in extracting informative posterior distributions when constrained by a limited number of forward simulations. Although sSVGD achieves better performance than SVGD and MH-McMC, it still lags behind the ADVI methods.

**APPENDIX F: Spurious Correlations in SVGD**

Suppose that we want to fit a Gaussian distribution in a three-dimensional space using only two samples, as shown in Figure F1. Say the true Gaussian distribution lies entirely within the X–Y plane. If we use point 1 and point 2, both located on this plane, the correlation between X and Y is represented by the blue line. This line captures the true linear relationship between X and Y accurately, without introducing any correlation with Z. However, if we use point 1 and point 3 instead, where point 3 lies off the X–Y plane, the correlation between these two samples is represented by the green line. When we project this green line onto the X–Y plane, it still aligns well with the true X–Y correlation. But because point 3 has a non-zero Z component, the line between points 1 and 3 introduces spurious correlations between X and Z, and between Y and Z. In reality, the true distribution has no correlation in the Z direction, and this would become apparent if we increase the number of samples of the true distribution: these samples would eventually indicate that there is equal probability of samples occurring above or below the X-Y plane, so there is no correlation in the Z direction.  The apparent correlations from the second 2-sample set are therefore artificial, and appear only because too few samples are available.

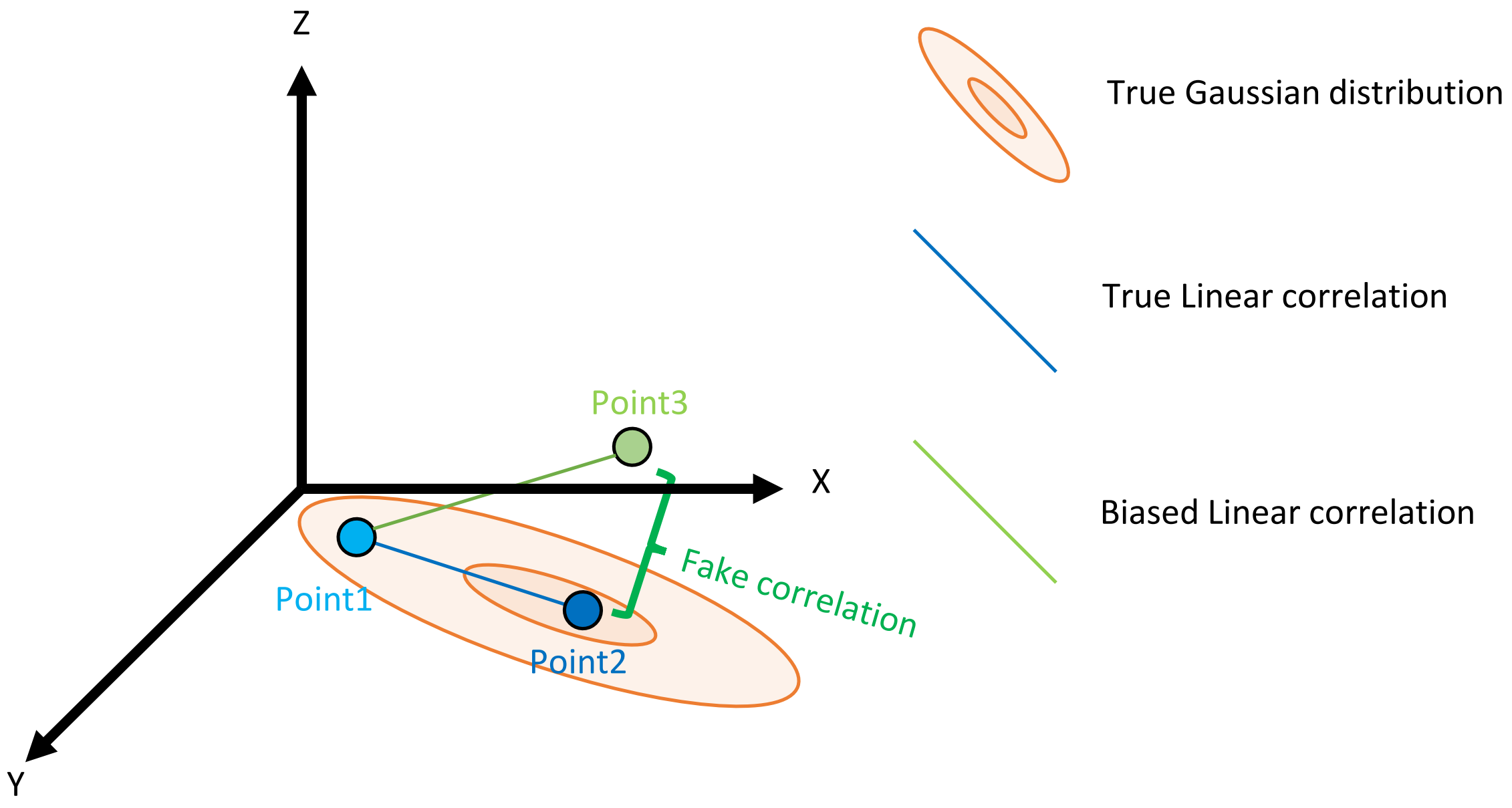

**Figure F1.** Spurious correlation caused by limited samples. We use two samples to fit a Gaussian distribution in a three-dimensional space, where the true correlation is only between parameters X and Y. Points 1 and 2 in the X-Y plane capture the true correlation between X and Y without introducing spurious correlation in Z. On the other hand, now assume that there is some (uncorrelated) Z variation in the samples, so that point 3 was sampled instead of point 2: while points 1 and 3 still reflect the correct X–Y correlation, they introduce spurious correlations with Z. This effect is due to the limited number of samples used: adding more samples would produce Z values above and below the X-Y plane with equal probability, cancelling the spurious Z correlation.

This illustrates a common problem when using a limited number of samples to approximate a high-dimensional posterior distribution. In our case, SVGD uses only 300 particles to approximate a posterior pdf in a 3,693-dimensional space. Since the number of samples is smaller than the number of dimensionalities, spurious correlations inevitably arise in the same manner as above. The unrealistic long-range correlation shown in Figure 16e, where a permeability cell appears correlated with another cell six rows away, is therefore likely to be caused by the limited number of samples afforded by SVGD. Note that we may also have observed this behaviour in the 2D case in Figure 6d, where SVGD produces a correlation matrix that includes correlations between permeability parameters located three rows apart while none of the other methods exhibit this behaviour. In principle we could mitigate this problem by increasing the number of particles, but this would increase the computational cost significantly.